\documentclass[11pt,onecolumn,draftcls]{IEEEtran}

\usepackage{amsmath,epsfig} 
\usepackage[english]{babel} 
\usepackage{amssymb}
\usepackage{color}
\usepackage{graphicx}
\usepackage{pgfrcs}
\usepackage{algpseudocode}
\usepackage{algorithm}
\usepackage{algorithmicx}
\usepackage{algcompatible}
\usepackage[noadjust]{cite}
\usepackage{caption}
\usepackage{multirow}

\begin{document} 

\title{A  Bayesian Consistent Dual Ensemble Kalman Filter for State-Parameter Estimation in Subsurface  Hydrology}

\author{Boujemaa~Ait-El-Fquih,~Mohamad~El~Gharamti, 
and~Ibrahim~Hoteit
\thanks{B. Ait-El-Fquih and I. Hoteit are with the 
Department of Earth Sciences and Engineering, 
King Abdullah University of Science and Technology (KAUST), 
Thuwal 23955-6900, Kingdom of Saudi Arabia  
(e-mail:boujemaa.aitelfquih@kaust.edu.sa, ibrahim.hoteit@kaust.edu.sa). M.E. Gharamti is with Mohn-Sverdrup Center for Global Ocean Studies and Operational Oceanography, 
Nansen Environmental and Remote Sensing Center (NERSC), 
Bergen 5006, Norway (e-mail:mohamad.gharamti@nersc.no)}}

\date{} 
\maketitle 
\begin{abstract}
Ensemble Kalman filtering (EnKF) is an efficient approach to addressing   uncertainties in subsurface groundwater models. The EnKF sequentially  integrates field data into simulation models to  obtain a better characterization of the model's state and parameters. These are generally  estimated  following joint and dual filtering strategies, in which, at each assimilation cycle, a  forecast step by the model is followed by an update step with incoming observations. The Joint-EnKF directly updates the augmented state-parameter  vector while the Dual-EnKF employs two separate filters, first estimating the parameters and then estimating the state based on the updated parameters. \textcolor{black}{ In this paper,  we reverse the order of the forecast-update steps following the one-step-ahead (OSA) smoothing formulation of the Bayesian filtering problem,  based on which we propose a new dual EnKF scheme; namely the Dual-EnKF$_{\rm OSA}$. Compared to the Dual-EnKF, this introduces a new update step to the state in a fully consistent Bayesian framework, which is shown to enhance the performance of the dual filtering approach without any significant increase in the computational cost}. Numerical experiments are conducted with a two-dimensional synthetic groundwater aquifer model. Assimilation experiments are performed under imperfect modeling conditions and various observational scenarios to assess the performance and robustness of the proposed Dual-EnKF$_{\rm OSA}$, and to evaluate its results against  those of the Joint-EnKF and Dual-EnKF. Simulation results suggest that the proposed Dual-EnKF$_{\rm OSA}$ scheme is able to successfully recover both the hydraulic head and the aquifer conductivity, further providing reliable estimates of 
their uncertainties. Compared with the standard Joint- and Dual-EnKFs, the proposed scheme is found more robust to different assimilation settings, such as the spatial and temporal distribution of the observations, and the level of noise in the data. Based on our experimental setups, it yields up to 25\% more accurate state and parameters estimates. 
\end{abstract}
%

\section{Introduction} 
In modern hydrology research, uncertainty quantification studies have focused on field applications, including surface and subsurface water flow, contaminant transport, and reservoir engineering. The motivations behind these studies were driven by the uncertain and stochastic nature of  hydrological systems. For instance, surface rainfall-runoff models that account for soil moisture and streamflows are subject to many uncertain parameters, such as free and tension water storage content, water depletion rates, and melting threshold temperatures \cite{Samuel2014}. Groundwater flow models, on the other hand, depend on our knowledge of  spatially variable aquifer properties, such as porosity and hydraulic conductivity, which are often poorly known \cite{Chen,Franssen2}. In addition, contaminant transport models that investigate the migration of pollutants in subsurface aquifers are quite sensitive to reaction parameters like sorption rates, radioactive decay, and biodegradation \cite{Gharamti2, Gharamti2014Hybrid}. To this end, it is important to study the variability of  hydrological parameters and reduce their associated uncertainties in order to obtain reliable simulations. To achieve this goal, hydrologists have used various inverse and Monte Carlo-based statistical  techniques with the standard procedure of pinpointing parameter values that, when integrated with the simulation models, allow some system-response variables (e.g., hydraulic head, solute concentration) to fit given observations \cite{Vrugt2003, Valstar2004, Alcolea2006, Feyen2007, Franssen1, Elsheikh2013}. Recently, sequential data assimilation techniques, such as \textcolor{black}{ the particle filter (PF), has been proposed   
to handle any type of statistical distribution, Gaussian or not, to properly deal with  strongly nonlinear systems  \cite{Chang}. The PF may require, however, a prohibitive number of particles (and thus model runs) to accurately sample the distribution of the state and parameters,   making this scheme computationally intensive for large-scale  hydrological applications \cite{livredoucet, Moradkhani, Montzka}. 
 This problem 
has been \textcolor{black}{partially addressed} 
by the popular ensemble Kalman filter (EnKF), which further provides  robustness, efficiency and non-intrusive formulation   \cite{Reichle2002, Vrugt2006, Zhou2011, Gharamti2013} to tackle the state-parameters estimation problem.}  

The EnKF is a 
  filtering technique that is relatively simple to implement, even with complex nonlinear models, requiring only an observation operator that maps the state variables from the model space into the observation space. 
Compared with traditional inverse {\color{black} and direct optimization techniques, which are generally based on least-squares-like formulations,} the EnKF has the advantage  of being able to account for model errors that are not only present in the uncertain parameters but also in the model structure and inputs, such as external forcings \cite{Franssen2}. In addition, and because of its  sequential  formulation, it does not require storing all past information about the states and parameters, leading to consequent savings in computational cost \cite{Mclaughlin2002, Gharamti2014Hybrid}. 

The EnKF is widely used in surface and subsurface hydrological studies to tackle state-parameters estimation problems. Two approaches are usually considered based on the joint and the dual state-parameter estimation strategies. The standard joint approach concurrently estimates the state and the parameters by augmenting the state variables with  the unknown parameters, that do not vary in time.  
The parameters could  also be set to follow an artificial evolution (random walk) before they get updated with incoming  observations \cite{Wan1999}. One of the early applications of the Joint-EnKF to subsurface groundwater flow models was carried in \cite{Chen}. In their study, a conceptual subsurface flow system was considered and ensemble filtering was performed to estimate the transient pressure field alongside the hydraulic conductivity. In a reservoir engineering application, \cite{Nevadal} considered a two-dimensional North Sea field model and considered the joint estimation problem of the dynamic pressure and saturation  on top of the static permeability field. Groundwater contamination problems were also tackled using the Joint-EnKF (e.g., \cite{Li2012,Gharamti2}), in which the hydraulic head,  contaminant concentration and spatially variable permeability and porosity parameters were estimated for  coupled groundwater flow and contaminant transport systems. 

In contrast with the joint approach, the dual approach runs two separate interactive EnKFs; one for the parameters and the other for the state. This separation is motivated  by the fact that it is the marginal posterior distributions, and not the joint posterior distribution of the state and parameters vector, that are the relevant quantities to be estimated  \cite{Moradkhani2005}.   
The Dual-EnKF has been applied to streamflow forecasting problems using rainfall-runoff models (e.g., \cite{Moradkhani2005,Samuel2014}). Recently, the authors in  \cite{Gharamti2014} successfully implemented the Dual-EnKF with a subsurface compositional flow model using chemical composite data and compared it to the standard Joint-EnKF. The authors concluded that the dual estimation scheme provides more accurate state and parameters estimates than  the joint scheme when implemented with  large enough ensembles. In terms of complexity, however, the dual scheme is computationally more  intensive,  
 as it requires integrating the filter ensemble twice with the numerical model at every  assimilation cycle. 

\textcolor{black}{In this study, we introduce a new Dual-EnKF algorithm   following the one-step-ahead (OSA) smoothing formulation of the Bayesian filtering problem,  in which the measurement-update step 
  precedes the forecast (or time-update) step. 
 Unlike the standard Dual-EnKF, this OSA smoothing-based filter, called Dual-EnKF$_{\rm OSA}$, is consistent with the Bayesian formulation of the state-parameter estimation problem and exploits the observations in both the state smoothing and forecast steps. It further  provides a theoretically sound framework for dual filtering and is shown to be beneficial in terms of estimation accuracy, as compared to the standard Joint- and Dual-EnKFs, without requiring any significant additional computational cost. } 
This is confirmed via synthetic numerical experiments using a groundwater flow model  and  estimating the hydraulic head and the conductivity parameter field. 

The rest of the paper is organized as follows. Section II reviews the standard Joint- and Dual-EnKF strategies. The Dual-EnKF$_{\rm OSA}$ is derived in Section III and its relation with the Joint- and Dual-EnKFs is discussed. Section IV presents a conceptual groundwater flow model and outlines the experimental setup. Numerical results are presented and discussed in Section V. Conclusions are offered in Section VI, followed by an Appendix.

\section{Joint and dual ensemble Kalman filtering} 
\subsection{Problem formulation}
Consider a discrete-time state-parameter dynamical system: 
\begin{equation} 
\label{DynSys} 
\left\{ 
\begin{array}{ccc} 
	\mathbf{x}_{n} &=& \mathcal{M}_{n-1}\left( \mathbf{x}_{n-1},\theta \right) + \eta_{n-1} \\
	\mathbf{y}_n &=& \mathbf{H}_n \mathbf{x}_n + \varepsilon_n  
\end{array}
\right. , 
\end{equation} 
in which $\mathbf{x}_n \in \mathbb{R}^{N_x}$ and $\mathbf{y}_n \in \mathbb{R}^{N_y}$ denote the system  state and  the observation at time $t_n$,  \textcolor{black}{ of dimensions $N_x$ and $N_y$  respectively}, and  $\theta \in \mathbb{R}^{N_{\theta}}$ is the parameter vector \textcolor{black}{of dimension $N_{\theta}$}. 
$\mathcal{M}_{n}$ is a nonlinear operator integrating the system state from  time  $t_{n}$ to $t_{n+1}$, 
\textcolor{black}{and the observational operator at time $t_{n}$, $\mathbf{H}_n$, is assumed to be linear for  simplicity; the proposed scheme is still valid in the  nonlinear case. }
The model process  noise, $\eta = {\{\eta_n\}}_{n \in \mathbb{N}}$, and the observation process noise,  $\varepsilon = {\{\varepsilon_n\}}_{n \in \mathbb{N}}$, are assumed to be independent, jointly independent and independent of ${\bf x}_0$ and  ${\theta}$, which, in turn, are assumed to be independent. Let also $\eta_n$ and $\varepsilon_n$ be Gaussian with zero means and  \textcolor{black}{covariances $\mathbf{Q}_n$ and   $\mathbf{R}_n$, respectively}. Throughout the paper, ${\bf y}_{0:n}  \stackrel{\rm def}{=} \{{\bf y}_0, {\bf y}_1, \cdots, {\bf y}_n\}$,  and $p({\bf x}_n)$ and $p({\bf x}_n|{\bf y}_{0:l})$ stand for  the \textcolor{black}{prior}  probability density  function (pdf) 
of ${\bf x}_n$ and the \textcolor{black}{posterior} pdf of ${\bf x}_n$ given ${\bf y}_{0:l}$, respectively. All other used pdfs are defined in a similar way. 

We focus on the state-parameter {\sl filtering} problem, say, the estimation at each time, $t_{n}$, of the state, ${\bf x}_n$, as well as the parameters vector,    $\theta$, from the history of the observations, ${\bf y}_{0:n}$. 
The standard solution of this problem is the {\sl a posteriori} mean (AM):  
\begin{eqnarray} 
\label{AM-estimate-x}
\mathbb{E}_{p\left( {\bf x}_n | {\bf y}_{0:n} \right)}\left[ {\bf x}_n \right]  & = & \int \! {\bf x}_n p\left( {\bf x}_n , {\theta} | {\bf y}_{0:n} \right) d{\bf x}_n d {\theta}, \\ 
\label{AM-estimate-theta} 
\mathbb{E}_{p\left( \theta | {\bf y}_{0:n} \right)}\left[ \theta \right] & = & \int \! \theta p\left( {\bf x}_n, \theta | {\bf y}_{0:n} \right) d{\bf x}_n d \theta , 
\end{eqnarray} 
which minimizes the {\sl a posteriori} mean square error. In practice, analytical  computation of (\ref{AM-estimate-x}) and (\ref{AM-estimate-theta}) is not feasible, mainly  due to the nonlinear character of the system. 
 The Joint- and   Dual-EnKFs have been introduced as efficient  schemes to compute approximations of (\ref{AM-estimate-x}) and (\ref{AM-estimate-theta}). Before reviewing these algorithms, 
 we recall the following classical results of stochastic sampling that we will extensively  use in the derivation of the filtering algorithms. 

\medskip \noindent
\textbf{Property 1} (Hierarchical sampling \cite{livreRobert-bayesianchoice-endEdit}).  
\label{proposition-hierarchical-sampling}
Assuming that one can sample from $p({\bf x}_1)$ and $p({\bf x}_{2}|{\bf x}_{1})$,  then a  sample,  ${\bf x}_2^{*}$, from $p({\bf x}_2)$ can be drawn as follows:     
\begin{enumerate} 
\item ${\bf x}_1^{*} \sim p({\bf x}_1)$; 
\item ${\bf x}_2^{*} \sim p({\bf x}_2|{\bf x}_1^{*})$. 
\end{enumerate}

\medskip \noindent
\textbf{Property 2} (Conditional sampling \cite{y-hoffman-et-al-1991}).
\label{proposition-conditional-sampling} 
Consider a Gaussian pdf, $p({\bf x} ,  {\bf y})$, with ${\bf P}_{xy}$ and ${\bf P}_{y}$ denoting  the cross-covariance of ${\bf x}$ and ${\bf y}$ and the covariance of ${\bf y}$, respectively. Then a sample, ${\bf x}^{*}$, from $p({\bf x}|{\bf y})$, can be drawn as follows:  
\begin{enumerate} 
\item $(\widetilde{\bf x} , \widetilde{\bf y}) \sim p({\bf x} , {\bf y})$;
\item ${\bf x}^{*} =  \widetilde{\bf x} + {\bf P}_{xy} {\bf P}_y^{-1} [{\bf y} - \widetilde{\bf y}]$.
\end{enumerate}

\subsection{The Joint- and Dual- EnKFs}
\subsubsection{The Joint-EnKF} 
\label{subsubsection-joint-enkf}
The key idea behind the Joint-EnKF is to transform the state-parameter system (\ref{DynSys})  into a classical state-space system based on the augmented state, ${\bf z}_n = \left[ {\bf x}_n^{T} \; \theta^T \right]^T$, on which the classical EnKF can  be  directly applied.  The new augmented state-space system can be written as: 
\begin{equation} 
\label{DynSys-augmented} 
\left\{ 
\begin{array}{ccc} 
	\mathbf{z}_{n} & = & \widetilde{\mathcal{M}}_{n-1}\left( \mathbf{z}_{n-1} \right) + \widetilde{\eta}_{n-1} \\
	\mathbf{y}_n & = & \widetilde{\mathbf{H}}_n \mathbf{z}_n + \varepsilon_n  
\end{array}
\right. , 
\end{equation}  
in which $\widetilde{\mathcal{M}}_{n-1}\left( \mathbf{z}_{n-1} \right) = \left[ 
\begin{array}{c}
\mathcal{M}_{n-1}({\bf z}_{n-1}) \\ 
 {\theta}
\end{array}
\right] $,  
$\widetilde{\eta}_{n-1} = \left[ {\eta}_{n-1}^T \; {\bf 0} \right]^T$, 
$\widetilde{\bf H}_n = \left[ {\bf H}_n \; {\bf 0} \right]$, with  ${\bf 0}$ a zero  matrix with appropriate dimensions. Starting at time $t_{n-1}$ from an analysis ensemble \textcolor{black}{of size $N_e$}, ${\{  {\bf x}_{n-1}^{a,(m)},{\theta}_{|n-1}^{(m)}  \}}_{m=1}^{N_{e}}$ representing $p({\bf z}_{n-1}| {\bf y}_{0:n-1})$, the EnKF uses the augmented state model ($1^{\rm st}$ Eq. of (\ref{DynSys-augmented})) to compute the forecast ensemble, ${\{  {\bf x}_{n}^{f,(m)},{\theta}_{|n-1}^{(m)}  \}}_{m=1}^{N_{e}}$, approximating $p({\bf z}_{n}| {\bf y}_{0:n-1})$. The observation model ($2^{\rm nd}$ Eq. of (\ref{DynSys-augmented})) is then used to obtain the analysis ensemble,  ${\{ {\bf x}_{n}^{a,(m)},{\theta}_{|n}^{(m)}  \}}_{m=1}^{N_{e}}$, at time  $t_n$. Let, for an ensemble ${\{ {\bf r}^{(m)}  \}}_{m=1}^{N_{e}}$, $\hat{\bf r}$ denotes its empirical mean and ${\bf S}_{\bf r}$  a  matrix with $N_e$-columns whose $m^{\rm th}$ column is defined as $\left({\bf r}^{(m)} - \hat{\bf r} \right)$. The Joint-EnKF steps can be summarized as follows: 
\begin{itemize} 
\item {\sl Forecast step:} The parameters vector members, ${\theta}_{|n-1}^{(m)}$, are kept invariant, while the state vector members, ${\bf x}_{n-1}^{a,(m)}$, are integrated in time through the dynamical model as:    
\begin{equation} 
\label{eq-forecast-JointEnKF}
{\bf x}_{n}^{f,(m)}  =  \mathcal{M}_{n-1}\left( {\bf x}_{n-1}^{a,(m)} ,  {\theta}_{|n-1}^{(m)} \right) + {\eta}_{n-1}^{(m)}; \;\;\; {\eta}_{n-1}^{(m)} \sim {\cal N}\left( {\bf 0} , {\bf Q}_{n-1} \right).   
\end{equation} 
An approximation of the forecast state, $\hat{\bf x}_{n|n-1}$, which is, by definition,  the mean of  $p({\bf x}_{n}|{\bf y}_{0:n-1})$ (similar to (\ref{AM-estimate-x})), is given by the empirical mean of the forecast ensemble, $\hat{\bf x}_n^f$. The associated forecast error covariance  is estimated as ${\bf P}_{{\bf x}_n^{f}} = \left( N_e-1\right)^{-1} {\bf S}_{{\bf x}_n^f} {\bf S}_{{\bf x}_n^f}^{T}$. 

\item {\sl Analysis step:} Once a new observation is available, all  members, ${\bf x}_{n}^{f,(m)}$ and ${\theta}_{|n-1}^{(m)}$, are updated as in the Kalman filter (KF): 
\begin{eqnarray}
\label{eq-JEnKF-yF}
{\bf y}_{n}^{f,(m)} & = & {\bf H}_n {\bf x}_n^{f,(m)}+{\varepsilon}_{n}^{(m)}; \;\;\; {\varepsilon}_{n}^{(m)} \sim {\cal N}({\bf 0} , {\bf R}_{n}), \\
\label{eq-JEnKF-xA} 
{\bf x}_{n}^{a,(m)}  & = & {\bf x}_{n}^{f,(m)} + {\bf P}_{ {\bf x}_{n}^f }  {\bf H}_{n}^T 
\underbrace{\left[ {\bf H}_n {\bf P}_{ {\bf x}_{n}^f }  {\bf H}_{n}^{T} + {\bf R}_n \right]^{-1} \left( {\bf y}_n - {\bf y}_n^{f,(m)} \right)}_{{\mu}_{n}^{(m)}}, \\ 
\label{eq-JEnKF-thetaA} 
{\theta}_{|n}^{(m)}  & = & {\theta}_{|n-1}^{(m)} + 
{\bf P}_{ \theta_{|n-1} ,{\bf y}_{n}^f}  \cdot {\mu}_{n}^{(m)},  
\end{eqnarray} 
with $ {\bf P}_{ \theta_{|n-1} ,{\bf y}_{n}^f} =  \left( N_e-1\right)^{-1}  {\bf S}_{{\theta}_{|n-1}} {\bf S}_{{\bf y}_n^f}^T$. The analysis estimates,  (\ref{AM-estimate-x}) and  (\ref{AM-estimate-theta}), and their error covariances, 
can thus be approximated by the analysis ensemble  means, $\hat{\bf x}_n^a$ and $\hat{{\theta}}_{|n}$, and  ${\bf P}_{{\bf x}_n^{a}} = \left( N_e-1\right)^{-1} {\bf S}_{{\bf x}_n^a} {\bf S}_{{\bf x}_n^a}^{T}$ and $ {\bf P}_{ \theta_{|n}} =  \left( N_e-1\right)^{-1}  {\bf S}_{{\theta}_{|n}} {\bf S}_{{\theta}_{|n}}^T$, respectively.  
\end{itemize}

\subsubsection{The Dual-EnKF} 
\label{subsec-dualEnKF} 
\textcolor{black}{ In contrast with the Joint-EnKF, 
 the Dual-EnKF is designed following a conditional estimation strategy, 
   operating as a succession of two EnKF-like filters. First, a (parameter) filter that computes ${\{{\theta}_{|n}^{(m)}\}}_{m=1}^{N_{e}}$ from ${\{ {\bf x}_{n-1}^{a,(m)} , {\theta}_{|n-1}^{(m)}\}}_{m=1}^{N_{e}}$,  is applied based on the following two steps.   
\begin{itemize} 
\item {\sl Forecast step:} The parameters  ensemble,  ${\{{\theta}_{|n-1}^{(m)}\}}_{m=1}^{N_{e}}$, is kept invariant, while the state samples are integrated in time as in (\ref{eq-forecast-JointEnKF}), to compute  the forecast ensemble, ${\{{\bf x}_{n}^{f,(m)}\}}_{m=1}^{N_{e}}$. 
\item {\sl Analysis step:} As in (\ref{eq-JEnKF-yF}), the observation forecast ensemble ${\{{\bf y}_{n}^{f,(m)}\}}_{m=1}^{N_{e}}$ is computed from  ${\{{\bf x}_{n}^{f,(m)}\}}_{m=1}^{N_{e}}$.  This is then used to update the parameters ensemble, ${\{{\theta}_{|n}^{(m)}\}}_{m=1}^{N_{e}}$, following  (\ref{eq-JEnKF-thetaA}). 
\end{itemize} 
Another   (state) filter is then applied to compute ${\{{\bf x}_{n}^{a, (m)}\}}_{m=1}^{N_{e}}$ from ${\{ {\bf x}_{n-1}^{a,(m)}\}}_{m=1}^{N_{e}}$ as well as the new parameter ensemble, ${\{{\theta}_{|n}^{(m)}\}}_{m=1}^{N_{e}}$, again in two steps that can be summarized as follows.  
\begin{itemize}
\item {\sl Forecast step:} Each member, ${\bf x}_{n-1}^{a,(m)}$, is propagated in time with  the dynamical model using the updated parameters ensemble:    
\begin{equation} 
\label{eq-forecast-DualEnKF}
\widetilde{\bf x}_{n}^{f,(m)}  =  \mathcal{M}_{n-1}\left( {\bf x}_{n-1}^{a,(m)} ,  {\theta}_{|n}^{(m)} \right).
\end{equation} 
\item {\sl Analysis step:} As in the parameter filter,  ${\{\widetilde{\bf x}_{n}^{f,(m)}\}}_{m=1}^{N_{e}}$ is used to compute ${\{\widetilde{\bf y}_{n}^{f,(m)}\}}_{m=1}^{N_{e}}$ as in (\ref{eq-JEnKF-yF}), which finally yields ${\{{\bf x}_{n}^{a,(m)}\}}_{m=1}^{N_{e}}$ as in   (\ref{eq-JEnKF-xA}). 
\end{itemize} }

\textcolor{black}{To better understand how the Dual-EnKF differs from the Joint-EnKF, we focus on how the outputs at  time $t_n$, namely, ${\bf x}_{n}^{a,(m)}$ and ${\theta}_{|n}^{(m)}$, are computed starting from the same ensemble members, ${\bf x}_{n-1}^{a,(m)}$ and ${\theta}_{|_{n-1}}^{(m)}$.  The  parameters members, ${\theta}_{|n}^{(m)}$,  are computed 
based on the same equation (\ref{eq-JEnKF-thetaA}) in both  algorithms. For the state members, ${\bf x}_{n}^{a,(m)}$, we have:
\begin{eqnarray} 
\label{update-xa--jenkf} 
{\bf x}_{n}^{a,(m)}  \!\!\! & \stackrel{\scriptsize \rm Joint-EnKF}{=} \!\!\! & \! {\cal M}_{n-1} \left( {\bf x}_{n-1}^{a,(m)} , \theta_{|n-1}^{(m)}\right) + {\bf P}_{ {\bf x}_{n}^f }  {\bf H}_{n}^T 
\underbrace{\left[ {\bf H}_n {\bf P}_{ {\bf x}_{n}^f }  {\bf H}_{n}^{T} + {\bf R}_n \right]^{-1} \left( {\bf y}_n - {\bf y}_n^{f,(m)} \right)}_{{\mu}_{n}^{(m)}}, \\
\label{update-xa--denkf} 
{\bf x}_{n}^{a,(m)}  \!\!\! & \stackrel{\scriptsize \rm Dual-EnKF}{=} \!\!\! & \underbrace{ {\cal M}_{n-1} ( {\bf x}_{n-1}^{a,(m)} , 
\overbrace{ 
\theta_{|n-1}^{(m)}  + 
{\bf P}_{ \theta_{|n-1} ,{\bf y}_{n}^f}  \cdot {\mu}_{n}^{(m)} 
}^{ \theta_{|n}^{(m)}  }
) 
}_{\widetilde{\bf x}_n^{f,(m)}} \nonumber  \\ 
& & + {\bf P}_{ \widetilde{\bf x}_{n}^f }  {\bf H}_{n}^T 
\underbrace{\left[ {\bf H}_n {\bf P}_{ \widetilde{\bf x}_{n}^f }  {\bf H}_{n}^{T} + {\bf R}_n \right]^{-1} \left( {\bf y}_n - \widetilde{\bf y}_n^{f,(m)} \right)}_{\widetilde{\mu}_{n}^{(m)}} . 
\end{eqnarray} 
Note that we ignore here the term for the process noise, ${\eta}_n$,  which is often the case in geophysics  applications. We further consider the same observation forecast members ${\bf  y}_{n}^{f,(m)}$ in both the Joint-EnKF (\ref{update-xa--jenkf})  and Dual-EnKF  (\ref{update-xa--denkf}) (which means that we consider the same measurement noise members, ${\varepsilon}_n^{(m)}$). As one can see, the Joint-EnKF  updates the state members  using one Kalman-like correction (term of $\mu_{n}^{(m)}$ in (\ref{update-xa--jenkf})), while  the Dual-EnKF  uses two  Kalman-like corrections. More specifically, the  Dual-EnKF updates first the parameters members, ${\theta}_{|n-1}^{(m)}$,  as in  the Joint-EnKF, leading to   ${\theta}_{|n}^{(m)}$; these  are then used to propagate    ${\bf x}_{n-1}^{a,(m)}$,  with the model  to provide  the ``forecast'' members  $\widetilde{\bf x}_{n}^{f,(m)}$.  
 The  $\widetilde{\bf x}_{n}^{f,(m)}$ are finally updated using a Kalman-like correction (term of $\widetilde{\mu}_{n}^{(m)}$ in (\ref{update-xa--denkf})),  
to obtain  the analysis members ${\bf x}_n^{a,(m)}$.  
Such a separation of the update steps was shown to provide more consistent estimates of the parameters, especially for strongly heterogeneous subsurface formations as suggested by \cite{WenChen2006} in their confirming-step EnKF algorithm. The dual-update framework was further shown to provide better performances than the Joint-EnKF at the cost of increased computational burden  (see for instance, \cite{Moradkhani2005,Samuel2014,Gharamti2014}).}

\subsubsection{Probabilistic formulation} 
\label{sec-proba-formulation}
Following a probabilistic formulation, the augmented state system (\ref{DynSys-augmented}) can be viewed as a continuous state Hidden Markov Chain (HMC) with transition density,   
\begin{equation} \label{eq-trans-MC-z}
p\left( {\bf z}_n|{\bf z}_{n-1} \right) = p\left( {\bf x}_n|{\bf x}_{n-1} , \theta \right)  p\left( \theta|\theta \right) = {\cal N}_{{\bf x}_n} \left( 
{\cal M}_{n-1} \left( {\bf x}_{n-1} , \theta \right) , {\bf Q}_{n-1} \right), 
\end{equation}
and likelihood, 
\begin{equation}
p\left( {\bf y}_n|{\bf z}_{n} \right)  = p\left( {\bf y}_n|{\bf x}_{n} \right) = {\cal N}_{{\bf y}_n} \left( {\bf H}_{n} {\bf x}_{n} , {\bf R}_n \right)   , 
\end{equation}
where ${\cal N}_{\bf v} ({\bf m}, {\bf C})$ represents a Gaussian pdf of argument ${\bf v}$ and parameters $({\bf m} , {\bf C})$. 

One can then easily verify that the Joint-EnKF  can be derived from a direct application of Properties 1 and 2 on the following classical generic formulas:   
\begin{eqnarray}
\label{eq-pp-markov-px-z-jointEnKF}
 p\left( {\bf z}_{n}|{\bf y}_{0:n-1} \right) & = & \int \! p\left( {\bf x}_{n}|{\bf x}_{n-1}, \theta \right) 
p\left( {\bf z}_{n-1}| {\bf y}_{0:n-1} \right) d{\bf x}_{n-1} , \\ 
\label{eq-pp-markov-py-jointEnKF}
 p\left( {\bf y}_{n}|{\bf y}_{0:n-1} \right) & = & \int p\left( {\bf y}_{n}|{\bf x}_{n} \right) 
p\left( {\bf x}_{n} | {\bf y}_{0:n-1} \right) d{\bf x}_{n} , \\ 
\label{eq-pp-bayes-px-z-jointEnKF}
p\left( {\bf z}_{n}|{\bf y}_{0:n} \right) & = & \frac{ p\left( {\bf z}_{n},{\bf y}_n|{\bf y}_{0:n-1} \right) }{ p\left( {\bf y}_{n}|{\bf y}_{0:n-1} \right)} .  
\end{eqnarray} 
\textcolor{black}{ Eq. (\ref{eq-pp-markov-px-z-jointEnKF}) refers to a {\sl Markovian} step (or time-update step) and uses the transition pdf, $p\left( {\bf x}_{n}|{\bf x}_{n-1}, \theta \right)$, of the Markov chain, ${\{{\bf z}_n \}}_n$, to compute the forecast pdf of ${\bf z}_n$ from the previous analysis pdf.  Eq. (\ref{eq-pp-bayes-px-z-jointEnKF}) refers to a {\sl Bayesian} step (or measurement-update step) since it uses   the Bayes' rule to update the forecast pdf of ${\bf z}_n$ using the current observation ${\bf y}_n$. Now,  establishing the link between the Joint-EnKF and Eqs. (\ref{eq-pp-markov-px-z-jointEnKF})-(\ref{eq-pp-bayes-px-z-jointEnKF}), one can show that} Property 1 and Eq. (\ref{eq-pp-markov-px-z-jointEnKF}) lead to the forecast ensemble of the state (\ref{eq-forecast-JointEnKF}). Property 1 and Eq. (\ref{eq-pp-markov-py-jointEnKF}) lead to the forecast ensemble of the observations (\ref{eq-JEnKF-yF}). Property 2 and Eq. (\ref{eq-pp-bayes-px-z-jointEnKF}) then provide the analysis ensemble of the state (\ref{eq-JEnKF-xA}) and the parameters (\ref{eq-JEnKF-thetaA}). 

Regarding the Dual-EnKF, the forecast ensemble of the state and  observations in the parameter filter can be obtained following the same process as in the Joint-EnKF. This is followed by the computation of the analysis ensemble of the parameters using Property 2 and 
\begin{equation}
\label{eq-pp-bayes-ptheta-dualEnKF}
p\left( \theta | {\bf y}_{0:n} \right) = \frac{ p\left( \theta ,{\bf y}_n|{\bf  y}_{0:n-1} \right)}{p\left( {\bf y}_{n}|{\bf y}_{0:n-1} \right)}. 
\end{equation} 
However, in the state filter, the ensemble, ${\{\widetilde{\bf x}_{n}^{f,(m)}\}}_{m=1}^{N_{e}}$,  obtained via Eq. (\ref{eq-forecast-DualEnKF}) in the forecast step  does not represent the  forecast pdf, $p({\bf x}_n | {\bf y}_{0:n-1})$, since Eq. (\ref{eq-forecast-DualEnKF})  involves ${\theta}_{|n}^{(m)}$ rather than ${\theta}_{|n-1}^{(m)}$.  Accordingly, the Dual-EnKF is basically a heuristic algorithm in spite of its proven  performance compared to the Joint-EnKF. 

In the next section, we follow  the same probabilistic framework to derive a new Dual-EnKF algorithm, in which the state is updated by the observations at both the measurement-update step and the time-update step, without violating the  Bayesian filtering formulation of the state-parameter estimation problem. Our goal is to build a fully Bayesian consistent dual-like EnKF that retains the separate formulation of the state and parameters update steps. 

\section{One-step-ahead smoothing-based Dual-EnKF (\textcolor{black}{Dual-EnKF$_{\rm OSA}$})}
The algorithm of this section is inspired by the fact that the classical \textcolor{black}{(time-update,  measurement-update) path  (\ref{eq-pp-markov-px-z-jointEnKF})-(\ref{eq-pp-bayes-px-z-jointEnKF})}, that  involves the forecast pdf $p({\bf z}_{n}|{\bf y}_{0:n-1})$ when moving from the analysis pdf $p({\bf z}_{n-1}|{\bf y}_{0:n-1})$ to the analysis pdf at the next time $p({\bf z}_{n}|{\bf y}_{0:n})$, is not the only possible one. Here,  \textcolor{black}{we reverse the order of the time-update and measurement-update steps}  to   derive a new dual algorithm  involving the one-step-ahead smoothing pdf, $p({\bf z}_{n-1}|{\bf y}_{0:n})$, between two successive analysis pdfs, $p({\bf z}_{n-1}|{\bf y}_{0:n-1})$ and $p({\bf z}_{n}|{\bf y}_{0:n})$. \textcolor{black}{This is  expected  to improve the filter estimates since  the state is constrained by more  observations than  the standard joint and dual schemes. 
  The reader may consult, for instance, \cite{desbouvries-et-al-2011}, in which the one-step-ahead smoothing-based filtering problem has been studied  and based on  which a class of KF- and PF-based algorithms have been derived. The more recent work in \cite{w-lee-and-c-farmer-2014}  also proposes a number of algorithms estimating both the system state and the model noise based on the same strategy.}  

\subsection{The one-step-ahead smoothing-based dual filtering algorithm} 
The analysis pdf, $p({\bf x}_{n} , \theta |{\bf y}_{0:n})$, can be computed from  $p({\bf x}_{n-1} , \theta | {\bf y}_{0:n-1})$ in two steps:
\begin{itemize}
\item {\sl Smoothing step:}  The  one-step-ahead smoothing pdf, $p({\bf x}_{n-1} , \theta |{\bf y}_{0:n})$, is first computed as, 
\begin{equation}
\label{eq-smoth-z-pp}
p({\bf x}_{n-1} , \theta |{\bf y}_{0:n}) \propto  \textcolor{black}{p({\bf y}_n | {\bf x}_{n-1}, \theta, {\bf y}_{0:n-1} )} p({\bf x}_{n-1} , \theta |{\bf y}_{0:n-1}) ,
\end{equation} 
\textcolor{black} {with,  
\begin{eqnarray}
\nonumber 
p({\bf y}_n | {\bf x}_{n-1}, \theta, {\bf y}_{0:n-1}) & = & \int p({\bf y}_n | {\bf x}_n, {\bf x}_{n-1} , \theta, {\bf y}_{0:n-1}) p({\bf x}_n | {\bf x}_{n-1} , \theta, {\bf y}_{0:n-1}) d{\bf x}_n,  \\
\label{eq-smoth-likliho-z}
 & = & \int p({\bf y}_n | {\bf x}_n) p({\bf x}_n | {\bf x}_{n-1} , \theta) d{\bf x}_n.
\end{eqnarray} 
Eq. (\ref{eq-smoth-likliho-z})  
 is derived from the fact that in the state-parameter model (\ref{DynSys}), the observation noise, $\varepsilon_n$,  and the model noise, $\eta_{n-1}$, are  independent of $({\bf x}_{n-1} , \theta)$ and  past observations ${\bf y}_{0:n-1}$. 
\newline 
The smoothing step (\ref{eq-smoth-z-pp}) is indeed  a  measurement-update step since   conditionally on  ${\bf y}_{0:n-1}$, Eq. (\ref{eq-smoth-z-pp}) translates the computation of the posterior, $p({\bf x}_{n-1}, \theta | {\bf y}_n)$, as a normalized product of the prior, $p({\bf x}_{n-1}, \theta)$, and the likelihood,    $p({\bf y}_n | {\bf x}_{n-1}, \theta)$ (recalling from (\ref{eq-smoth-likliho-z}) that $p({\bf y}_n | {\bf x}_{n-1}, \theta, {\bf y}_{0:n-1})$ $= p({\bf y}_n | {\bf x}_{n-1}, \theta)$). }
\item {\sl Forecast step:}  The smoothing pdf at  $t_{n-1}$ is then used to  compute the current analysis pdf, $p({\bf x}_{n} , \theta |{\bf y}_{0:n})$, as 
\begin{equation}
\label{eq-smoth-transition-pp-x}
p({\bf x}_n, \theta | {\bf y}_{0:n}) = \int \textcolor{black}{p({\bf x}_n | {\bf x}_{n-1} , \theta , {\bf y}_{0:n})} p({\bf x}_{n-1} , \theta | {\bf y}_{0:n}) d{\bf x}_{n-1}, 
\end{equation}
\textcolor{black}{ with,  
\begin{equation}
\label{eq-smoth-a-posteriori-transition-pp-x}
p({\bf x}_n | {\bf x}_{n-1} , \theta , {\bf y}_{0:n}) \propto p({\bf y}_n | {\bf x}_n) p({\bf x}_{n} | {\bf x}_{n-1} , \theta ), 
\end{equation} 
which, in turn, arises from the fact that $\varepsilon_n$ and $\eta_{n-1}$  are  independent of $({\bf x}_{n-1}, \theta)$ and ${\bf y}_{0:n-1}$ (see smoothing step above). 
  We note here that only the (marginal) analysis pdf of ${\bf x}_n$, $p({\bf x}_{n}  |{\bf y}_{0:n})$, is of interest  since the analysis pdf of $\theta$ has  already been computed in the smoothing step. 
\newline 
From  (\ref{eq-smoth-a-posteriori-transition-pp-x}),  $p({\bf x}_n | {\bf x}_{n-1} , \theta , {\bf y}_{0:n}) = p({\bf x}_n | {\bf x}_{n-1} , \theta , {\bf y}_{n})$.  Thereby, there is a similarity between  Eq. (\ref{eq-smoth-transition-pp-x}) and the forecast step  (\ref{eq-pp-markov-px-z-jointEnKF}), in the sense that (\ref{eq-smoth-transition-pp-x}) can be seen as a forecast step once the current observation ${\bf y}_n$ is known, {\sl i.e.}, (\ref{eq-smoth-transition-pp-x}) coincides with ``(\ref{eq-pp-markov-px-z-jointEnKF}) given the observation ${\bf y}_n$''. }
\textcolor{black}{Accordingly, and without abuse of language, we refer to   (\ref{eq-smoth-transition-pp-x})-(\ref{eq-smoth-a-posteriori-transition-pp-x}) as the  {\sl forecast} step.}  
\end{itemize} 

\subsection{Ensemble Formulation}
Since it is not possible to derive the analytical  solution of (\ref{eq-smoth-z-pp})-(\ref{eq-smoth-a-posteriori-transition-pp-x}) because of the nonlinear character of the model, ${\cal M}(.)$, we propose an EnKF-like formulation, \textcolor{black}{assuming that $p({\bf y}_n, {\bf z}_{n-1} | {\bf y}_{0:n-1})$ is Gaussian for all $n$. This assumption implies that $p({\bf z}_{n-1} | {\bf y}_{0:n-1})$, $p({\bf z}_{n-1} | {\bf y}_{0:n})$ and $p({\bf y}_{n} | {\bf y}_{0:n-1})$ are Gaussian.}

\subsubsection{Smoothing step} 
Starting at time $t_{n-1}$, from an analysis ensemble, ${\{ {\bf x}_{n-1}^{a,(m)} , \theta_{|_{n-1}}^{(m)} \}}_{m=1}^{N_e}$, 
one can use Property 1 in Eq. (\ref{eq-smoth-likliho-z}) to sample the observation forecast ensemble, ${\{ {\bf y}_{n}^{f,(m)} \}}_{m=1}^{N_e}$,  as  
\begin{eqnarray} 
\label{smooth-enkf-algo-ynf-1-0} 
\textcolor{black}{ {\bf x}_{n}^{f,(m)} }  & \textcolor{black}{=} &  \textcolor{black}{ {\cal M}_{n-1}({\bf x}_{n-1}^{a,(m)} , \theta_{|n-1}^{(m)}) + {\eta}_{n-1}^{(m)} ; } \\   
\label{smooth-enkf-algo-ynf-1} 
{\bf y}_{n}^{f,(m)}  & = & {\bf H}_n  \textcolor{black}{ {\bf x}_{n}^{f,(m)} } 
 +  {\varepsilon}_n^{(m)},  
\end{eqnarray}
with ${\eta}_{n-1}^{(m)} \sim {\cal N}({\bf 0} , {\bf Q}_{n-1})$ and ${\varepsilon}_n^{(m)} \sim {\cal N}({\bf 0} , {\bf R}_n)$.  Property 2 is then used in Eq. (\ref{eq-smoth-z-pp}) to compute the smoothing ensemble, ${\{ {\bf x}_{n-1}^{s,(m)} , \theta_{|_{n}}^{(m)} \}}_{m=1}^{N_e}$, as  
\begin{eqnarray}
\label{smooth-enkf-algo-x}
{\bf x}_{n-1}^{s,(m)}   & =  &  {\bf x}_{n-1}^{a,(m)} + {\bf P}_{{\bf x}_{n-1}^{a} , {\bf y}_{n}^{f}} 
\underbrace{  {\bf P}_{{\bf y}_n^f}^{-1} \left( {\bf y}_n - {\bf y}_{n}^{f,(m)} \right) }_{\nu_{n}^{(m)}},  \\
\label{smooth-enkf-algo-theta}
\theta_{|n}^{(m)}  & = &  \theta_{|n-1}^{(m)} + {\bf P}_{\theta_{|n-1} , {\bf y}_{n}^{f}}  
\cdot \nu_{n}^{(m)} . 
\end{eqnarray} 
The (cross-) covariances in equations (\ref{smooth-enkf-algo-x}) and (\ref{smooth-enkf-algo-theta}) are defined and practically evaluated from the ensembles as in the EnKF (see Subsection \ref{subsubsection-joint-enkf}): 
\begin{eqnarray} 
\label{equat-cross-smooth-1}
{\bf P}_{{\bf x}_{n-1}^{a} , {\bf y}_{n}^{f}} &=& \left( N_e-1 \right)^{-1} {\bf S}_{{\bf x}_{n-1}^a} {\bf S}_{{\bf y}_{n}^{f}}^T, \\
{\bf P}_{{\bf y}_{n}^{f}} &=& \left( N_e-1 \right)^{-1} {\bf S}_{{\bf y}_{n}^{f}} {\bf S}_{{\bf y}_{n}^{f}}^T, \\
\label{equat-cross-smooth-3}
{\bf P}_{ \theta_{|n-1} , {\bf y}_{n}^{f} } &=& \left( N_e-1 \right)^{-1} {\bf S}_{{\theta}_{|n-1}} {\bf S}_{{\bf y}_n^f}^T.  
\end{eqnarray} 

\subsubsection{Forecast step} 
\label{forecast-pratical-deriv-subsection}
 The analysis ensemble, ${\{ {\bf x}_{n}^{a,(m)}\}}_{m=1}^{N_e}$, can be obtained from  ${\{ {\bf x}_{n-1}^{s,(m)} , \theta_{|_{n}}^{(m)} \}}_{m=1}^{N_e}$ using Property 1 in Eq. (\ref{eq-smoth-transition-pp-x}), once the {\sl a posteriori} transition pdf, $p({\bf x}_n | {\bf x}_{n-1} , \theta , {\bf y}_{n})$, is computed via Eq. (\ref{eq-smoth-a-posteriori-transition-pp-x}). Furthermore, one can verify that  Eq. (\ref{eq-smoth-a-posteriori-transition-pp-x}) leads to a Gaussian pdf: 
\begin{equation}
\label{a-posteriori-transtition-pdf}
p({\bf x}_n | {\bf x}_{n-1} , \theta , {\bf y}_{n}) = 
{\cal N}_{{\bf x}_n} 
\left( 
{\cal M}_{n-1}({\bf x}_{n-1} , \theta) + \widetilde{\bf K}_n 
\left( 
{\bf y}_n - {\bf H}_n {\cal M}_{n-1}({\bf x}_{n-1} , \theta) 
\right) 
, 
\widetilde{\bf Q}_{n-1} 
\right), 
\end{equation} 
with 
$
\widetilde{\bf K}_n  = {\bf Q}_{n-1} {\bf H}_n^T 
\left[ {\bf H}_n {\bf Q}_{n-1} {\bf H}_n^T + {\bf R}_n \right]^{-1} 
$
and 
$\widetilde{\bf Q}_{n-1} = {\bf Q}_{n-1} - \widetilde{\bf K}_n {\bf H}_n {\bf Q}_{n-1}$. However, when the state dimension, $N_x$, is very large, the computational cost of $\widetilde{\bf K}_n$ and $\widetilde{\bf Q}_{n-1}$ (which may be an off-diagonal matrix even when ${\bf Q}_{n-1}$ is diagonal) may become  prohibitive. 
\textcolor{black}{One way to avoid this problem is to directly sample from $ p({\bf x}_n | {\bf x}_{n-1} ,  \theta , {\bf y}_{n})$ without   explicit computation of this pdf in (\ref{a-posteriori-transtition-pdf}). Let ${\{ \widetilde{\bf x}_n^{(m)} ({\bf x}_{n-1} , \theta) \}}_{m=1}^{N_e}$ denotes an ensemble of samples drawn from $ p({\bf x}_n | {\bf x}_{n-1} ,  \theta , {\bf y}_{n})$. The notation $\widetilde{\bf x}_n^{(m)} ({\bf x}_{n-1} , \theta)$ refers to a function $\widetilde{\bf x}_n^{(m)}$ of $({\bf x}_{n-1} , \theta)$;  similar notations hold for  $\widetilde{\xi}_n^{(m)}(.)$ and $ \widetilde{\bf y}_n^{(m)}(.)$ in (\ref{sample-a-poetserior-pdf-given-x-1}) and (\ref{sample-a-poetserior-pdf-given-x-2}), respectively. Using Properties 1 and 2, an explicit form of such samples can be obtained as (see Appendix), 
\begin{eqnarray}
\label{sample-a-poetserior-pdf-given-x-1} 
\widetilde{\xi}_n^{(m)} ({\bf x}_{n-1} , \theta) & = & {\cal M}_{n-1}  ({\bf x}_{n-1} , \theta) + {\eta}_{n-1}^{(m)};\;\;\; {\eta}_{n-1}^{(m)} \sim {\cal N} ({\bf 0} , {\bf Q}_{n-1}), \\
\label{sample-a-poetserior-pdf-given-x-2} 
\widetilde{\bf y}_n^{(m)} ({\bf x}_{n-1} , \theta) & = & {\bf H}_n \widetilde{\xi}_n^{(m)} ({\bf x}_{n-1} , \theta) + {\varepsilon}_n^{(m)};\;\;\;  {\varepsilon}_{n}^{(m)} \sim {\cal N} ({\bf 0} , {\bf R}_{n}), \\ 
\label{sample-a-poetserior-pdf-given-x-3} 
\widetilde{\bf x}_n^{(m)} ({\bf x}_{n-1} , \theta) & = & 
\widetilde{\xi}_n^{(m)} ({\bf x}_{n-1} , \theta) 
+ {\bf P}_{\widetilde{\xi}_n , \widetilde{\bf y}_n} 
{\bf P}_{\widetilde{\bf y}_n}^{-1} \left[ 
{\bf y}_n - \widetilde{\bf y}_{n}^{(m)}({\bf x}_{n-1} , \theta)
\right],  
\end{eqnarray} 
where the (cross)-covariances, $ {\bf P}_{\widetilde{\xi}_n , \widetilde{\bf y}_n}$ and ${\bf P}_{\widetilde{\bf y}_n}$, are evaluated from  ${\{  \widetilde{\xi}_n^{(m)} ({\bf x}_{n-1} , \theta)  \} }_{m=1}^{N_e}$ and ${\{  \widetilde{\bf y}_n^{(m)} ({\bf x}_{n-1} , \theta)  \} }_{m=1}^{N_e}$, similarly to (\ref{equat-cross-smooth-1})-(\ref{equat-cross-smooth-3}). 
Now, using Property 1 in Eq. (\ref{eq-smoth-transition-pp-x}), one can compute an analysis  ensemble, ${\{ {\bf x}_{n}^{a,(m)}\}}_{m=1}^{N_e}$, from the smoothing ensemble,   ${\{ {\bf x}_{n-1}^{s,(m)} , \theta_{|_{n}}^{(m)} \}}_{m=1}^{N_e}$,  using the functional form (\ref{sample-a-poetserior-pdf-given-x-3}). More precisely, we obtain, ${\bf x}_{n}^{a,(m)} = \widetilde{\bf x}_{n}^{(m)}({\bf x}_{n-1}^{s,(m)} , \theta_{|n}^{(m)})$, which is equivalent to set ${\bf x}_{n-1} = {\bf x}_{n-1}^{s,(m)}$ and $\theta = \theta_{|n}^{(m)}$ in  (\ref{sample-a-poetserior-pdf-given-x-1})-(\ref{sample-a-poetserior-pdf-given-x-3}). 
 }

\subsubsection{Summary of the algorithm} 
\label{subsection-algo-vb-dual-enkf-summary}
Starting  from an analysis ensemble, ${\{ {\bf x}_{n-1}^{a,(m)} , \theta_{|n-1}^{(m)} \}}_{m=1}^{N_e}$, at time $t_{n-1}$, the updated ensemble of both the state and parameters at time $t_{n}$ is obtained with the following two steps:
\begin{itemize}
\item {\sl Smoothing step:} \textcolor{black}{ The  state forecast ensemble,  ${\{ {\bf x}_{n}^{f,(m)}  \}}_{m=1}^{N_e}$, is first computed by (\ref{smooth-enkf-algo-ynf-1-0}), and then used to compute the  observation  forecast ensemble,  ${\{ {\bf y}_{n}^{f,(m)}  \}}_{m=1}^{N_e}$, as in (\ref{smooth-enkf-algo-ynf-1}). This latter is then used to compute the one-step-ahead smoothing ensemble of the state, ${\{ {\bf x}_{n-1}^{s,(m)} \}}_{m=1}^{N_e}$,  and parameters, ${\{  \theta_{|_{n}}^{(m)} \}}_{m=1}^{N_e}$, following Eqs. (\ref{smooth-enkf-algo-x}) and (\ref{smooth-enkf-algo-theta}), respectively. }
\item {\sl Forecast step:} The analysis ensemble of the state ${\{ {\bf x}_{n}^{a,(m)} \}}_{m=1}^{N_e}$ is  obtained as:
\begin{eqnarray}
 \xi_{n}^{(m)}  & = & {\cal M}_{n-1} \left( {\bf x}_{n-1}^{s, (m)} , \theta_{|n}^{(m)} \right) \textcolor{black}{ + {\eta}_{n-1}^{(m)};  \;\;\; {\eta}_{n-1}^{(m)} \sim {\cal N}({\bf 0} , {\bf Q}_{n-1}), } \\
\widetilde{\bf y}_{n}^{f,(m)} & = & {\bf H}_n \xi_{n}^{(m)} + {\varepsilon}_n^{(m)}; \;\;\; {\varepsilon}_n^{(m)} \sim {\cal N}({\bf 0} , {\bf R}_n), \\ 
\label{smooth-enkf-algo-x-bis} 
{\bf x}_{n}^{a,(m)}  & =  &  \xi_{n}^{(m)} + {\bf P}_{\xi_{n} , \widetilde{\bf y}_{n}^{f}} 
  {\bf P}_{\widetilde{\bf y}_n^f}^{-1} ({\bf y}_n - \widetilde{\bf y}_{n}^{f,(m)}) ,  
\end{eqnarray} 
with $ {\bf P}_{{\xi}_{n} , \widetilde{\bf y}_{n}^{f}} = \left( N_e-1 \right)^{-1} {\bf S}_{{\xi}_{n}} {\bf S}_{\widetilde{\bf y}_{n}^{f}}^T $  and  
$ {\bf P}_{\widetilde{\bf y}_{n}^{f}} = \left( N_e-1\right)^{-1} {\bf S}_{\widetilde{\bf y}_{n}^{f}} {\bf S}_{\widetilde{\bf y}_{n}^{f}}^T $.  
\end{itemize} 

\textcolor{black}{ In contrast with the Dual-EnKF, which uses $\theta_{|n}^{(m)}$ and ${\bf x}_{n-1}^{a,(m)}$ for computing ${\bf x}_{n}^{a,(m)}$ (see Eq. (\ref{update-xa--denkf})),  the proposed Dual-EnKF$_{\rm OSA}$ uses $\theta_{|n}^{(m)}$ and the smoothed  state members, ${\bf x}_{n-1}^{s,(m)}$, which are the ${\bf x}_{n-1}^{a,(m)}$ after an update  with the current observation, ${\bf y}_n$, following (\ref{smooth-enkf-algo-x}).} Therefore, when including the Kalman-like correction term as well, the observation, ${\bf y}_n$,  is used three times in the the Dual-EnKF$_{\rm OSA}$ in a fully consistent Bayesian formulation, but only twice in the Dual-EnKF.  \textcolor{black}{ This means that   the Dual-EnKF$_{\rm OSA}$ exploits the observations more efficiently  than the Dual-EnKF, which should provide more information for improved state and parameters estimates}. 

\textcolor{black}{ 
Despite the smoothing framework of the Dual-EnKF$_{\rm OSA}$, this algorithm obviously   addresses the  state forecast problem as well. As discussed in the smoothing step above, the (one-step-ahead) forecast members are inherently computed. 
%
The $j$-step-ahead forecast member, denoted  by ${\bf x}_{n+ j|n}^{(m)}$ for  $j \geq2 $, can be computed following a recursive procedure, where   for $\ell =  2, 3, \cdots, j$, one has    
\begin{equation}
\label{eq-forcast-j-step}
{\bf x}_{n+\ell|n}^{(m)} = {\cal M}_{n+\ell-1} ({\bf x}_{n+\ell-1|n}^{(m)} , \theta_{|n}^{(m)}) + {\eta}_{n+\ell-1}^{(m)}, \quad {\eta}_{n+\ell-1}^{(m)} \sim {\cal N} ({\bf 0} , {\bf Q}_{n+\ell-1}) .
\end{equation} 
%
}

\subsection{Complexity of the Joint-EnKF,  Dual-EnKF, and Dual-EnKF$_{\rm OSA}$} 
The computational complexity of the different state-parameter EnKF schemes can be split between the forecast (time-update) step and the analysis (measurement-update) step. \textcolor{black}{ The Joint-EnKF requires  $N_e$ model runs (for forecasting the state ensemble) and $N_e$ Kalman corrections (for updating the forecast ensemble). This is practically doubled when using the Dual-EnKF, since the latter requires $2 N_e$ model runs and $2N_e$ Kalman corrections; $N_e$ corrections for each of the forecast state ensemble  and the forecast parameter ensemble. As presented in the previous section, the Dual-EnKF$_{\rm OSA}$ smoothes  the state estimate  at the previous time step before updating the parameters and the state at the current time. Thus, the Dual-EnKF$_{\rm OSA}$ requires as many model runs ($2 N_e$) as the Dual-EnKF, and an additional $N_e$ correction steps to apply  smoothing.  In large scale geophysical applications, the correction step of the ensemble members is often computationally not significant   compared to the cost of integrating the model in the forecast step. The approximate computational complexity and memory storage for each algorithm are summarized in Table I. The tabulated complexities for each method are valid under the assumption that $N_y \ll N_x$, {\sl i.e.},  the number of state variables is much larger than the number of observations. This is generally the case for subsurface flow applications due to budget constraints given the consequent  costs needed for preparing, drilling, and maintaining subsurface wells. }

\begin{table}[h]
\caption{\small Approximate computational complexities of the Joint-EnKF, the Dual-EnKF, and the Dual-EnKF$_{\rm OSA}$ algorithms. Notations are as follows. $N_x$: number of state variables, $N_\theta$: number of parameter variables, $N_y$: number of observations, $N$: number of assimilation cycles, $N_e$: ensemble size, $\mathcal{C}_x$: state model cost ($=N_x^2$ is the linear KF), $\mathcal{C}_{\theta}$: parameter model cost (usually free $\equiv$ identity), $\mathcal{C}_y$: observation operator cost ($=N_yN_x$ in the linear KF), $\mathcal{S}_x$: storage volume for one state vector, $\mathcal{S}_\theta$: storage volume for one parameter vector.}
\vspace{-2mm}
\begin{center}
\begin{tabular}{ l c c c } \hline
\textbf{Algorithm}         	& \textbf{Time-update} & \textbf{Measurement-update} & \textbf{Storage} \\  \noalign{\smallskip} \hline {\smallskip}
\textit{Joint-EnKF}        	& $NN_e\left( \mathcal{C}_x + \mathcal{C}_{\theta} \right)$   	& $NN_e\left( \mathcal{C}_y + N_yN_\theta \right) + NN_e^2\left( N_x+N_\theta \right)$ 	& $2NN_e \left( \mathcal{S}_x + \mathcal{S}_\theta \right)$ \\
\textit{Dual-EnKF}        	& $NN_e\left( 2\mathcal{C}_x + \mathcal{C}_{\theta} \right)$ 	& $2NN_e\mathcal{C}_y + NN_e^2\left( N_x+N_\theta  \right)$ 					& $2NN_e \left( \mathcal{S}_x + \mathcal{S}_\theta \right)$  \\ 
\textit{Dual-EnKF$_{\rm OSA}$} 	& $NN_e\left( 2\mathcal{C}_x + \mathcal{C}_{\theta} \right)$ 	& $2NN_e\mathcal{C}_y + NN_e^2\left( 2N_x+N_\theta  \right)$ 					& $2NN_e \left( \mathcal{S}_x + \mathcal{S}_\theta \right)$  \\ \hline
\end{tabular}
\end{center}
\end{table}

\section{Numerical experiments} 
\label{num-sect-label}
\subsection{Transient groundwater flow problem}
We adopt in this study the subsurface flow problem of \cite{Bailey2010}. The system consists of a  two-dimensional (2D) transient flow with an areal aquifer area of 0.5 km$^2$ (Figure 1). Constant head boundaries of 20 m and 15 m are placed on the west and east ends of the aquifer, respectively, with an average saturated thickness, $b$, of $25$ m. The north and south boundaries are assumed to be Neumann with no flow conditions (Figure 1). The mesh is discretized using a cell-centered finite difference scheme with $10$ m $\times$ $20$ m rectangles, resulting in $2500$ elements. The following 2D saturated groundwater flow system is solved:
\begin{equation}
\frac{\partial}{\partial x}\left( T_x \frac{\partial h}{\partial x} \right) + \frac{\partial}{\partial y}\left( T_y \frac{\partial h}{\partial y} \right) = S\frac{\partial h}{\partial t} + q,
\end{equation}
where $T$ is the transmissivity [L$^2$T$^{-1}$], which is related to the conductivity, $K$, through $T = Kb$, $h$ is the hydraulic head [L], $t$ is time [T], $S$ is storativity [-], and $q$ denotes the sources as recharge or sinks due to pumping wells [LT$^{-1}$]. Unconfined aquifer conditions are simulated by setting $S = 0.20$ to represent the specific yield. A log-conductivity field is generated using the sequential Gaussian simulation toolbox, GCOSIM3D \cite{Gomez1993}, with a mean of $-13$ log(m/s), a variance of $1.5$ log(m$^2$/s$^2$), and a Gaussian variogram with a range equal to $250$ m in the x-direction and $500$ m in the y-direction (Figure 1). 

\begin{figure}[ht]
\centering \vskip -.32cm
\includegraphics[width=10cm]{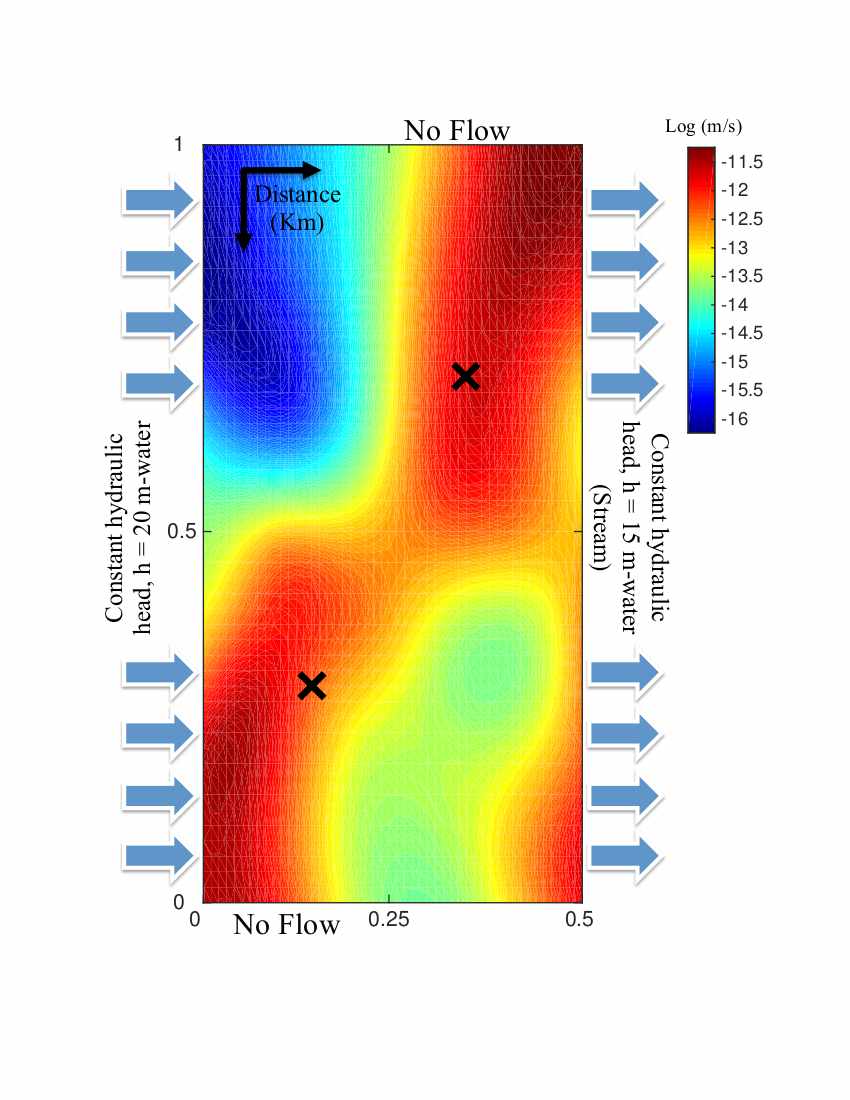} \vskip -2cm
\caption{\small  Plan view of the conceptual model for the 2D transient groundwater flow problem. East and west boundaries are Dirichlet with a given prescribed hydraulic heads. North and south boundaries are impermeable (no flow boundaries). The reference log-conductivity field obtained using the sequential Gaussian simulation code \cite{Gomez1993}. A Gaussian variogram model is considered with a mean of -13 $\log$(m/s), variance of 1.5 $\log$(m$^2$/s$^2$), and range equal to 250 m and 500 m in the x and y directions, respectively.}
\end{figure}

\textcolor{black}{We consider a dynamically complex experimental setup that is  similar to a real-world application and is based on various time-dependent external forcings. The recharge is assumed spatially heterogenous and sampled using the GCOSIM3D toolbox \cite{Gomez1993} with statistical parameters shown in Table II. Three different pumping wells (PW) are inserted within the aquifer domain and can be seen in the right panel of Figure 2 (square symbols). From these wells, transient pumping of groundwater takes place with different daily values as plotted in the left panel of Figure 2. The highest pumping rates are associated with PW2 with an average daily rate of $5.935 \times 10^{-7}$ m. Smaller temporal variations in water pumping rates are assigned to PW1 and PW3. Three other monitoring wells (MW1, MW2, MW3) are also placed within the aquifer domain to evaluate the groundwater flow filters estimates.  We further assess the prediction skill of the model after data assimilation using a control well (CW) placed in the middle of the aquifer (indicated by a diamond symbol).}

Prior to assimilation, a reference run is first conducted for each experimental setup using the prescribed parameters above, and is considered as the truth. \textcolor{black}{We simulate the groundwater flow system over a year-and-a-half period using the classical fourth-order Runge Kutta method with a time step of $12$ hours.} The initial hydraulic head configuration \textcolor{black}{is obtained after a 2-years model spin-up starting from a uniform $15$ m head.} Reference heterogenous recharge rates are used in the setup as explained before. The water head changes (in m-water) \textcolor{black}{after $18$ months are displayed with contour lines in the left panel of Figure 3.} One can notice larger variations in the water head in the lower left corner of the aquifer domain, consistent with the high conductivity values in that region. The effects of transient pumping in addition to the heterogenous recharge rates \textcolor{black}{are also well observed in the vicinity of the pumping wells.}

\begin{figure}[h]
\centering \vskip -0.05cm 
\includegraphics[width=12cm]{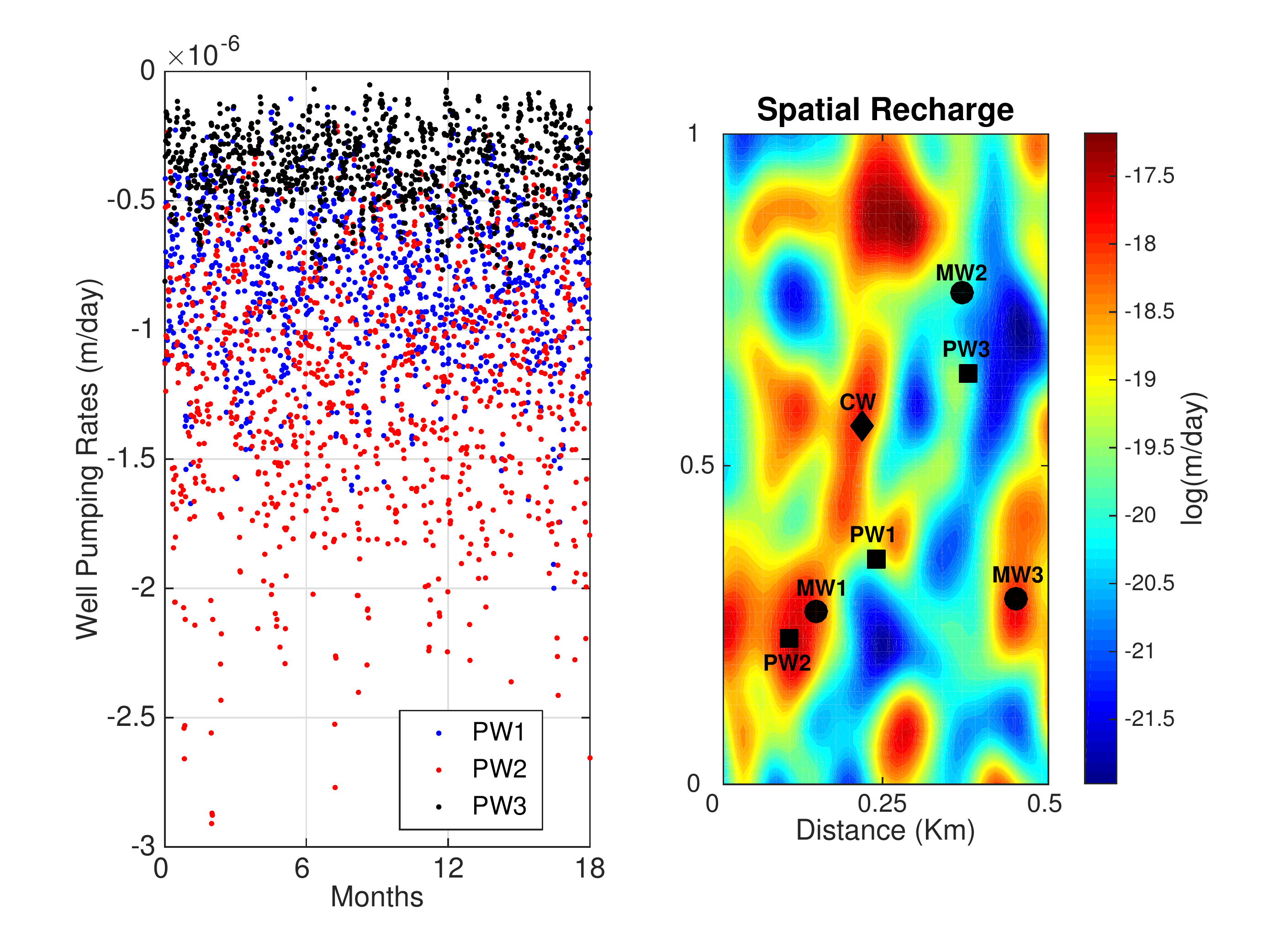} \vskip -.5cm
\caption{\small Left-Panel: Daily transient reference pumping rates from wells PW1, PW2 and PW3. Negative values indicate pumping or groundwater that is being removed from the aquifer. Right-Panel: Reference heterogenous spatial recharge values obtained using the sequential Gaussian simulation code \cite{Gomez1993} with parameters given in Table II. The black squares represent the pumping wells whereas the black circles denote the position of 3 monitoring wells. The black diamond is a control well.}
\end{figure}

\begin{figure}[h]
\centering \vskip - .15cm 
\includegraphics[width=13cm]{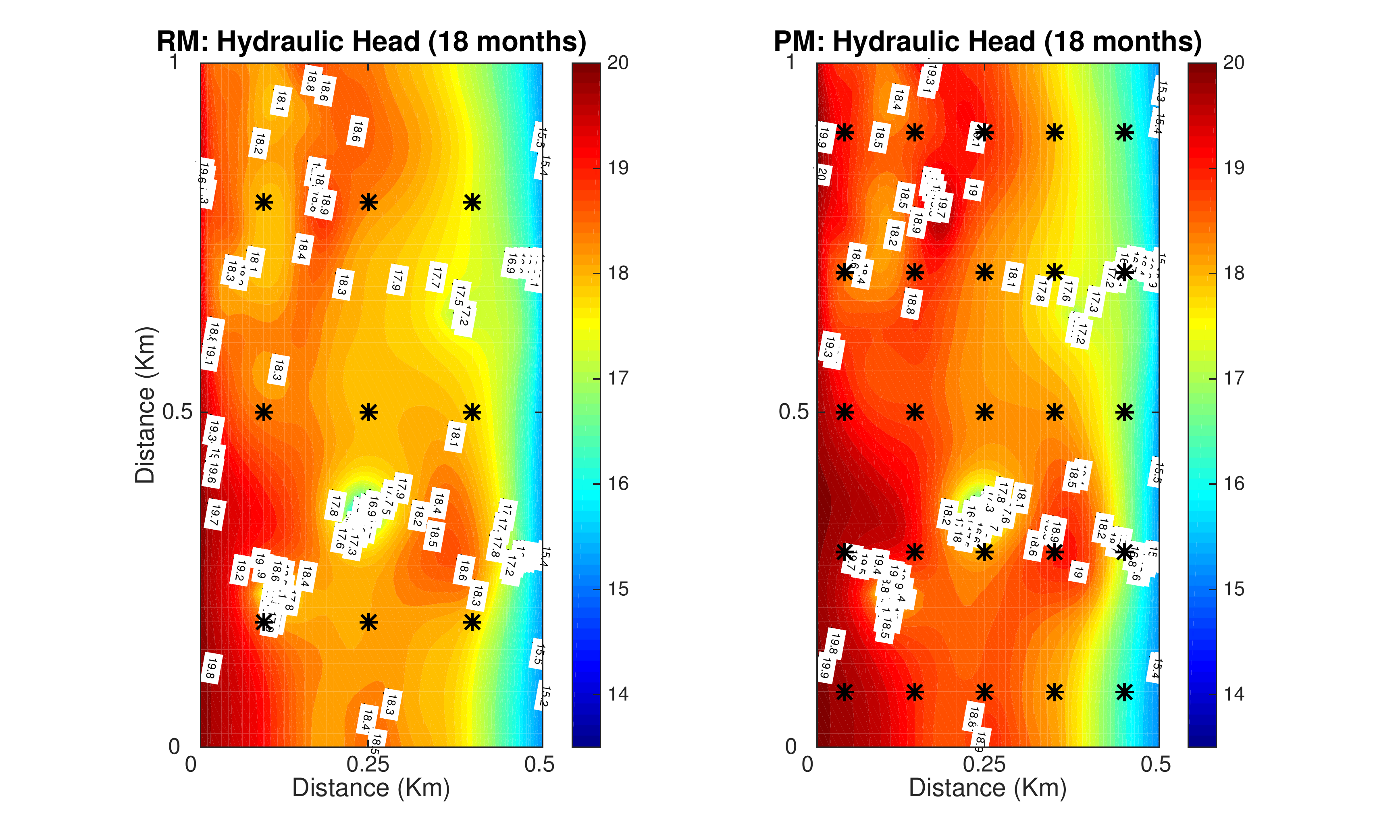} \vskip -0.1cm
\caption{\small Groundwater flow contour maps obtained using the reference run (left panel) and the perturbed forecast model (right panel) after $18$ months of simulation. The well locations from which head data are extracted are shown by black asterisks. In the left panel, we show the first network consisting of nine wells. In the right panel, the other network with 25 wells is displayed.}
\end{figure}

\subsection{Assimilation Experiments}
To imitate a realistic setting, we impose various perturbations on the reference model and  set our goal to estimate the water head and the hydraulic conductivity fields using an imperfect forecast model and perturbed data extracted from the reference (true) run. This experimental framework is  known as ``twin-experiments". In the forecast model, we perturb both transient pumping and spatial recharge rates. The perturbed recharge field, as compared to the reference recharge in Figure 2, is sampled with different variogram parameters as shown in Table II. \textcolor{black}{Pumping rates from PW1, PW2 and PW3  are perturbed by adding a Gaussian noise with mean zero and standard deviation equal to $20\%$ of the reference transient rates. The flow field simulated by  the forecast (perturbed) model after $18$ months is shown in the right panel of Figure 3. Compared to the reference field, there are clear spatial differences in the hydraulic head, especially around the first and second pumping wells.} 

\begin{table} \vskip -.2cm
\caption{\small Parameters of the random functions for modeling the spatial distributions of the reference and perturbed recharge fields. The ranges in $x$ and $y$ directions for the variorum model are given by $\lambda_x$ and $\lambda_y$, respectively. $\tau$ denotes the rotation angle of one clockwise rotation around the positive y-axis.}
\vspace{-2mm}
\begin{center}
\begin{tabular}{ccccccc} \hline
\textbf{Recharge} & \textbf{Mean} & \textbf{Variance} & \textbf{Variogram} & $\lambda_x$ & $\lambda_y$ & $\tau$ \\ \hline
Reference Field & -20 (m/day) & 1.03 (m/day)$^2$ & Gaussian & $50$ (m) & $100$ (m) & $45^{\circ}$  \\
Perturbed Field & -20 (m/day) & 1.21 (m/day)$^2$ & Gaussian & $50$ (m) & $50$ (m) & $45^{\circ}$ \\
	\hline
	\end{tabular}
\end{center}	
\end{table}

To demonstrate the effectiveness of the proposed Dual-EnKF$_{\rm OSA}$ compared to the standard Joint- and Dual-EnKFs, we evaluate the filter's accuracy and robustness under different experimental scenarios. We conduct a number of sensitivity experiments,  changing: (1) the ensemble size, (2) the temporal frequency of available observations, (3) the number of observation wells in the domain, and (4) the measurement error. For the frequency of the observations, we consider $6$ scenarios in which hydraulic head measurements are extracted from the reference run \textcolor{black}{every 1, 3, 5, 10, 15, and 30 days. Of course,  ${\bf x}_n^{a,(m)}$ is equal to  ${\bf x}_n^{f,(m)}$ when no observation is assimilated.} We also test \textcolor{black}{four different observational networks  assuming  9, 15, 25 and 81 wells uniformly distributed throughout the aquifer domain (Figure 3 displays two of these networks; with $9$ and $25$ wells).} We evaluate the  algorithms \textcolor{black}{under $9$ different scenarios in which the observations were perturbed with Gaussian noise of zero mean and a standard deviation equal to 0.10, 0.15, 0.20, 0.25, 0.30, 0.50, 1, 2 and 3 m.} 

To initialize the filters, we follow \cite{Gharamti2014} and perform a 5-years simulation run using the perturbed forecast model starting from the mean hydraulic head of the reference run solution. Then, we randomly select a set of $N_e$ hydraulic head snapshots to form the initial state ensemble. By doing so, the dynamic head changes that may occur  in the aquifer are well represented by the initial ensemble. The corresponding parameters' realizations are sampled with the geostatistical software, GCOSIM3D, using the same variogram parameters of the reference conductivity field but conditioned on two hard measurements as indicated by black crosses in Figure 1. The two data points capture some parts of the high conductivity regions in the domain, and thus one should expect a poor representation of the low conductivity areas in the initial $\log(K)$ ensemble. This is a challenging case for the filters especially when  a sparse observational network is considered. To ensure consistency between the hydraulic heads and the conductivities at the beginning of the assimilation, we conduct a spin-up of the whole state-parameters ensemble for a $6$-months period using perturbed recharge time-series for each ensemble member.  

The filter estimates resulting from the different filters are evaluated based on their average absolute forecast errors (AAE) and their average ensemble spread (AESP):
\begin{eqnarray}
AAE &=& N_x^{-1}N_e^{-1}\sum_{j=1}^{N_e}\sum_{i=1}^{N_x} \bigg| \mathbf{x}^{f,e}_{j,i} - \mathbf{x}_{i}^t \bigg|, \\
AESP &=& N_x^{-1}N_e^{-1}\sum_{j=1}^{N_e}\sum_{i=1}^{N_x}\bigg| \mathbf{x}^{f,e}_{j,i} - \hat{\mathbf{x}}^{f,e}_{i} \bigg| ,
\end{eqnarray}
where $\mathbf{x}_{i}^t$ is the reference ``true" value of the variable at cell $i$, $\mathbf{x}^{f,e}_{j,i}$ is the forecast ensemble value of the variable, and $\hat{\mathbf{x}}^{f,e}_i$ is the forecast ensemble mean at location $i$. 
  AAE measures the estimate-truth misfit and AESP measures the ensemble spread, or the confidence in the estimated values \cite{Franssen2}. We further assess the accuracy of the estimates by plotting the resulting field and variance maps of both hydraulic head and conductivities. 

\section{Results and Discussion}
\subsection{Sensitivity to the ensemble size}
We first study the sensitivity of the three algorithms to the ensemble size, $N_e$. In realistic groundwater applications, we would be restricted to small ensembles due to  computational limitations. Obtaining accurate state and parameter estimates with small ensembles is thus desirable. \textcolor{black}{We carry the experiments using three ensemble sizes, $N_e =$  50, 100 and 300, and we fix the frequency of the  observations to half a day, the number of wells to nine (Figure 3, left observation network) and the measurement error to $0.50$ m.} We plot the resulting AAE time series  of \textcolor{black}{the state and parameters in Figure 4.} As shown, the performance of the Joint-EnKF, Dual-EnKF and Dual-EnKF$_{\rm OSA}$ improves as the ensemble size increases, \textcolor{black}{reaching a mean AAE of $0.161$, $0.160$, and $0.156$ m-water for $N_e = 300$,} respectively. The Joint-EnKF and the Dual-EnKF exhibit similar behaviors, with a slight advantage for the Dual-EnKF. As demonstrated in \cite{Gharamti2014}, the Dual-EnKF is generally expected to produce more accurate results  only when large enough ensembles are used. We have tested the Joint- and the Dual-EnKFs using 1000 members and found that the Dual EnKF is around $9\%$  more accurate in term of AAE.
  The proposed Dual-EnKF$_{\rm OSA}$ is the most accurate in all tested scenarios. On average, with changing ensemble size, the Dual-EnKF$_{\rm OSA}$ leads to about $7\%$ improvement compared with  the joint and dual schemes. 
\textcolor{black}{In terms of the conductivity estimates, the proposed scheme  produces more accurate estimates for all three ensemble sizes. At the early assimilation stage, the three schemes seem to provide similar results,  but this eventually changes after $6$ months and the Dual-EnKF$_{\rm OSA}$ clearly outperforms the standard schemes as the complexity of the model increases because of the perturbed recharge and pumping rates.} 

\begin{figure}[h]
\centering \vskip .00025cm 
\includegraphics[width=15cm]{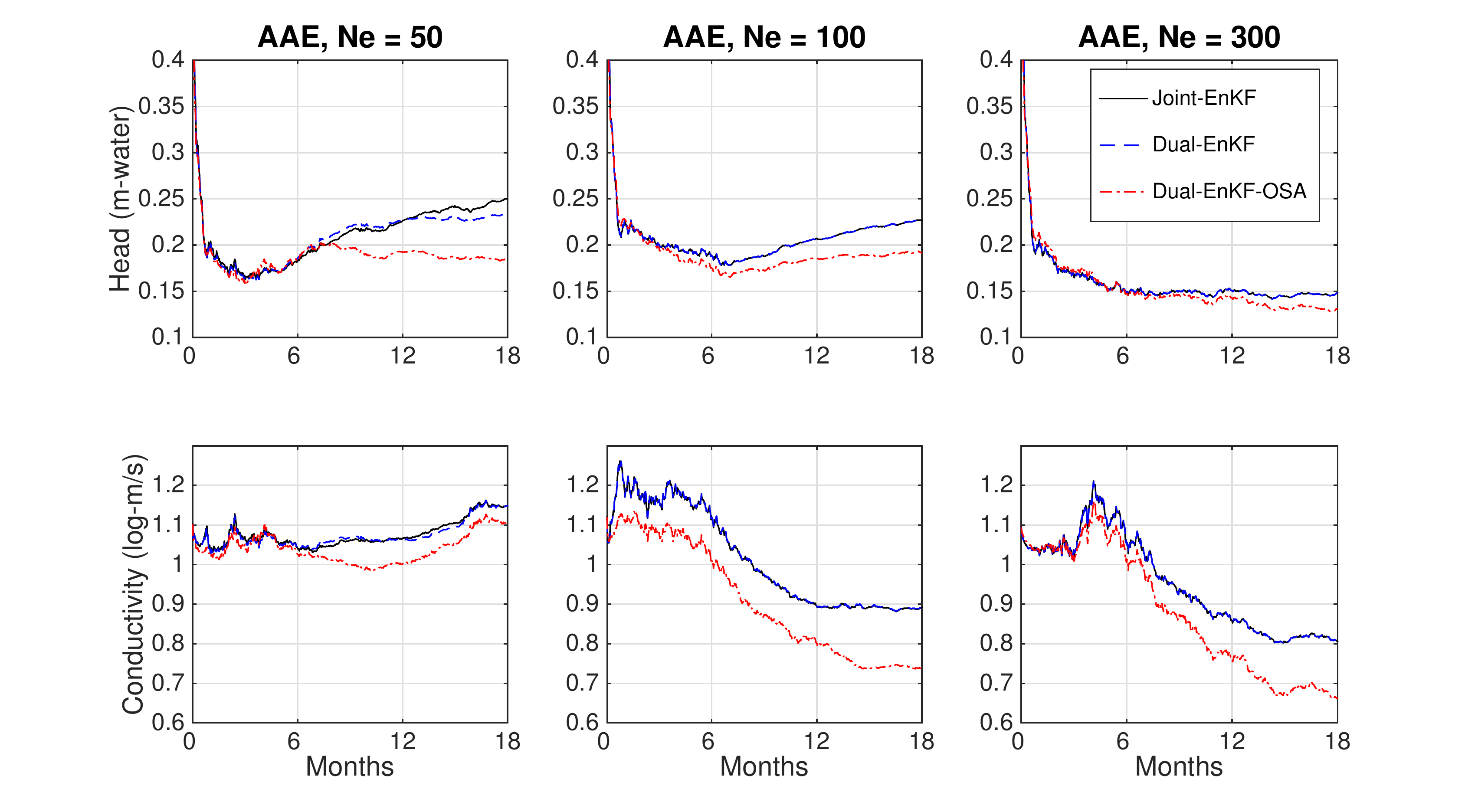} \vskip -0.1cm
\caption{\small AAE time-series of the hydraulic head and conductivity using the Joint-EnKF, Dual-EnKF and Dual-EnKF$_{\rm OSA}$. Results are shown for $3$ scenarios in which assimilation of hydraulic head data are obtained from nine wells every $0.5$ days. The three experimental scenarios use $50$, $100$ and $300$ ensemble members with 0.50 m as the measurement error.}
\end{figure}

\textcolor{black}{We furthermore examined the uncertainties of these forecast estimates by computing  the spread of both the hydraulic head and conductivity ensembles. To do this, we evaluated the mean AESP of both variables and tabulated the results for the three ensemble sizes in Table III. For all schemes and as expected, the spread seems to increase as the ensemble size increases. Compared to the joint and  the dual schemes, the Dual-EnKF$_{\rm OSA}$ retains the smallest mean AESP for all cases, suggesting  more confidence in the  head and conductivity estimates.}  

\begin{table}[h] \vskip -.2cm
\caption{\small Mean average ensemble spread (AES) of the water head and the hydraulic conductivity for three different ensemble sizes. The reported values are given for the Joint-EnKF, Dual-EnKF and the proposed Dual-EnKF$_{\rm OSA}$.}
\vspace{-2mm}
\begin{center}
\begin{tabular}{ccccc|cccc} \hline
 & & \textbf{Concentration} & & & & & \textbf{Conductivity} & \\ \cline{2-9}
 & $N_e=50$ & $N_e=100$ & $N_e=300$ & & & $N_e=50$ & $N_e=100$ & $N_e=300$ \\ \cline{1-9}
\textit{Joint-EnKF} & 0.12294 & 0.144 & 0.20014 & & & 1.0763 & 1.0144 & 0.95128  \\
\textit{Dual-EnKF} & 0.1256 & 0.14469  & 0.20081 & & & 1.0745 & 1.0155 & 0.95129 \\
\textit{Dual-EnKF$_{\rm OSA}$} & 0.11737 & 0.14125 & 0.18259 & & & 1.0388 & 0.90654 & 0.8791 \\
	\hline
	\end{tabular}
\end{center}	
\end{table}

In terms of computational cost, we note that our assimilation results were obtained using a $2.30$ GHz MAC workstation and $4$ cores for parallel looping while integrating the ensemble members. The Joint-EnKF is the least  intensive  requiring $70.61$ sec to perform a year-and-a-half assimilation run using $50$ members. The Dual-EnKF and Dual-EnKF$_{\rm OSA}$, on the other hand, require $75.37$ and $77.04$ sec, respectively. The Dual-EnKF is more intensive than the Joint-EnKF because it includes an additional propagation step of the ensemble members as discussed in Section III.C. Likewise, the proposed Dual-EnKF$_{\rm OSA}$ requires both an additional propagation step and an update step of the state members. \textcolor{black}{Its computational complexity is thus greater than the joint scheme and roughly equivalent to that of the Dual-EnKF. Note that in the current setup the cost of integrating the groundwater model is not very significant as compared to the cost of the update step. This might however not be the case in large-scale hydrological applications.}   

\subsection{Sensitivity to the frequency of observations}
In the second set of experiments, we change only the temporal frequency of observations, i.e., the times at which head observations are assimilated. We implement  the three filters with $100$ members and use data from nine observation wells perturbed with 0.10 m noise. 

Figure 5 plots the mean AAE of the hydraulic conductivity estimated using the three filters for the six different observation frequency scenarios. The Dual- and Joint-EnKFs lead to comparable performances, but the latter performs slightly better when data are assimilated more frequently, \textcolor{black}{i.e., every five and three days.} The performance of the proposed Dual-EnKF$_{\rm OSA}$, as seen from the plot, is rather effective and results in more accurate estimates than those obtained by the other two filters. The largest improvements resulting from this scheme are obtained when assimilating data every $1$, $3$, and $5$ days. The improvements over the joint and the dual schemes decrease as the frequency of observations in time decreases. The reason for this is related to the nature of the Dual-EnKF$_{\rm OSA}$ algorithm, which adds a one-step-ahead-smoothing to the analyzed head ensemble members before updating the forecast parameters and the state samples. Therefore, the more data available in time, the greater the number of smoothing steps applied, and hence the better the characterization of the state and parameters. To illustrate, the smoothing step of the state ensemble enhances its statistics and eventually provides more consistent state-parameters cross-correlations to better predict the data at the next update step. \textcolor{black}{When assimilating hydraulic head data on a daily basis, the proposed Dual-EnKF$_{\rm OSA}$ leads to about 24\% more accurate conductivity estimates than the Joint and Dual-EnKFs.}  

\begin{figure}[h]
\centering \vskip .0025cm 
\includegraphics[width=10cm]{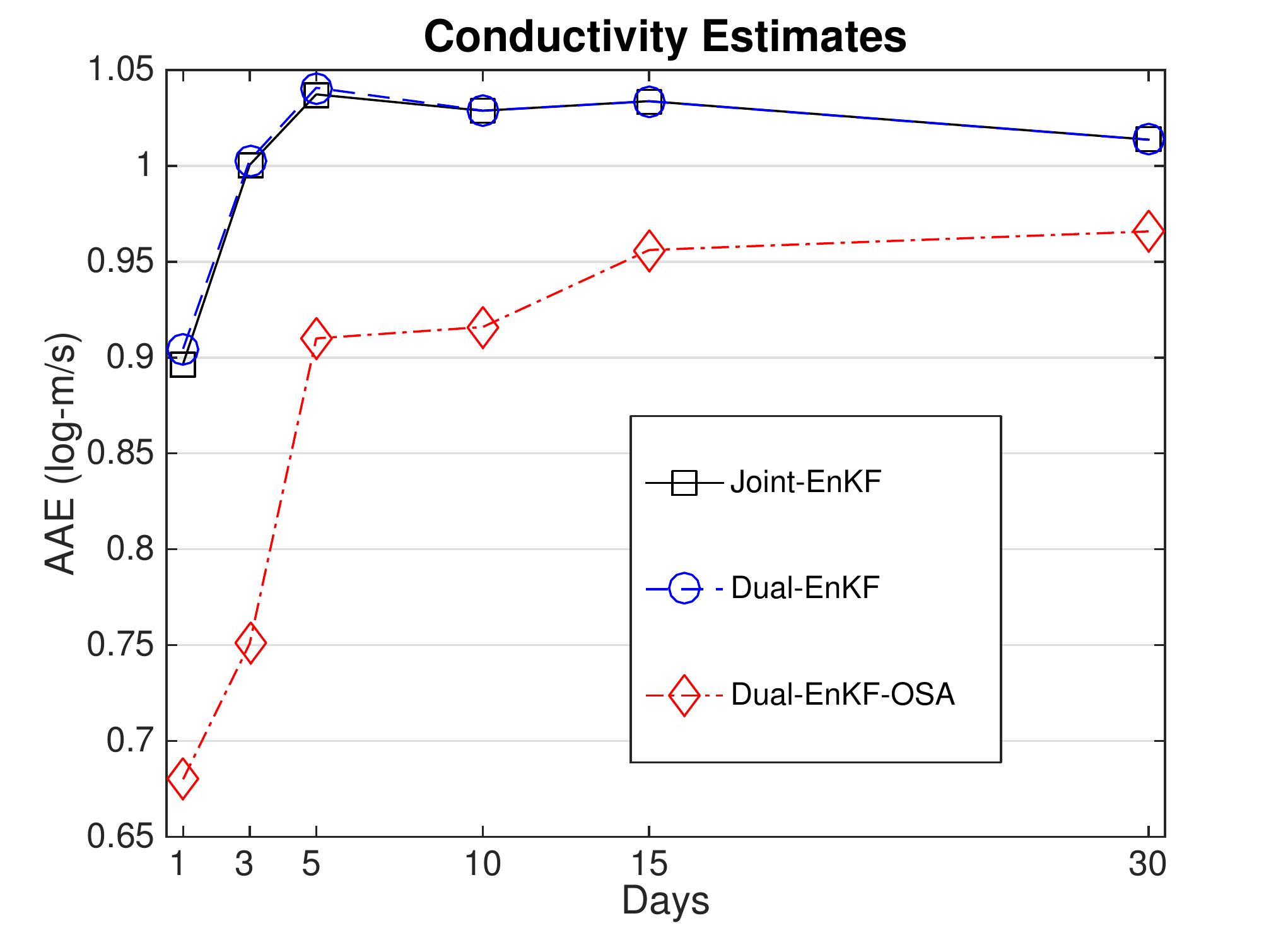} \vskip -0.1cm
\caption{\small Mean average absolute errors (AAE) of log-hydraulic conductivity, $\log(K)$, obtained using the Joint-EnKF, Dual-EnKF, and Dual-EnKF$_{\rm OSA}$ schemes. Results are shown for $6$ different scenarios in which assimilation of hydraulic head data are obtained from nine wells every $1$, $3$, $5$, $10$, $15$ and $30$ days. All $6$ experimental scenarios use $100$ ensemble members and $0.10$ m as the measurement error.}
\end{figure}

\textcolor{black}{We also compared the hydraulic head estimates when changing the temporal frequency of observations.} Similar to the parameters, the improvements of the Dual-EnKF$_{\rm OSA}$ algorithm over the other schemes become significant when more data are assimilated over time. Overall, the benefits of the proposed scheme are  more pronounced for the estimation of the parameters because the conductivity values at all aquifer cells are indirectly updated using hydraulic head data, requiring more observations for efficient estimation. 

\textcolor{black}{One effective way to evaluate the estimates of the state is to examine the evolution of the reference heads and the forecast ensemble members at various aquifer locations. For this, we plot in Figure 6 the true and the estimated time-series change in hydraulic head at the assigned monitoring wells as they result from the Joint-EnKF, Dual-EnKF and the Dual-EnKF$_{\rm OSA}$. We use $100$ ensemble members and assume the 9 data points are available every five days. At MW1, the performance of the three filters is quite similar and they all successfully reduce the uncertainties and recover the true evolution of the hydraulic head at that location. We note that between the $5^{\rm th}$ and the $9^{\rm th}$ month, the Dual-EnKF seems to underestimate the reference values of the hydraulic head as compared to the other two schemes. At MW2 and MW3, the ensemble spread of all three filters shrinks shortly after the assimilation starts but remains larger than those at MW1. The proposed Dual-EnKF$_{\rm OSA}$ efficiently recovers the reference trajectory of MW2 and MW3. The ensemble head values obtained using the Joint- and the Dual-EnKFs at MW2 are less accurate. Furthermore, the Joint and the Dual-EnKF ensemble members tend to underestimate the reference hydraulic head at MW3 over the first $6$ months of assimilation. Beyond this, there is a clear overestimation of the head values, especially by the Dual-EnKF, up to the end of the first year.}  

\begin{figure}[h]
\centering 
\includegraphics[width=14cm]{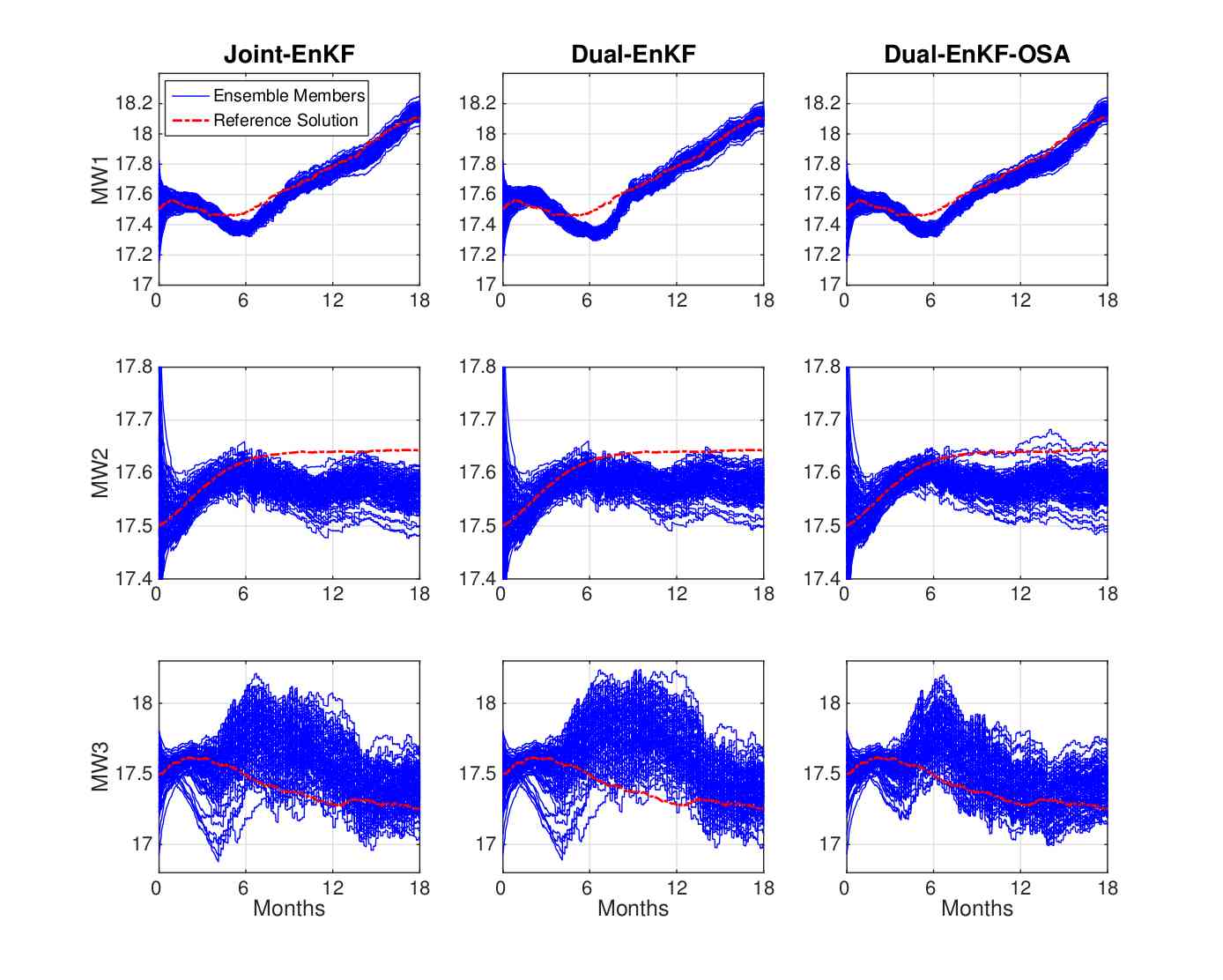} \vskip -0.5cm
\caption{\small Reference (dashed) and predicted (solid) hydraulic head evolution at monitoring wells MW1, MW2, and MW3. Results are obtained using the Joint-EnKF and the Dual-EnKF$_{\rm OSA}$ schemes with $100$ members, $5$ days as observation frequency, $9$ observation wells, and $0.10$ m of measurement noise.}
\end{figure}


\subsection{Sensitivity to the number of observations}
We further study the robustness of the proposed Dual-EnKF$_{\rm OSA}$ against the Joint- and Dual-EnKFs to different numbers of observation wells inside the aquifer domain. We thus compare our earlier estimates resulting from only nine wells, five days observation frequency, and 0.10 m measurement error with a new set of estimates \textcolor{black}{resulting from more dense networks with 15, 25 and 81 wells.} 
  \textcolor{black}{Figure 7 plots the time-series curves of the AAE as they result from the four observational scenarios for hydraulic head and conductivity. As shown, the behavior of the filters improves as more data are assimilated.} Clearly, the proposed scheme provides the best estimates over the entire simulation window. More precisely, \textcolor{black}{and towards the end of assimilation,} the Dual-EnKF$_{\rm OSA}$ with only nine data points exhibits less \textcolor{black}{forecast errors for conductivity than does the Dual-EnKF with 81 data points.}  Likewise \textcolor{black}{when assimilating head data from 15 and 25 wells,} the proposed algorithm outperforms the Dual-EnKF and yields more accurate hydraulic head estimates by the end of the simulation time.

\begin{figure}[h]
\centering \vskip .0025cm 
\includegraphics[width=13cm]{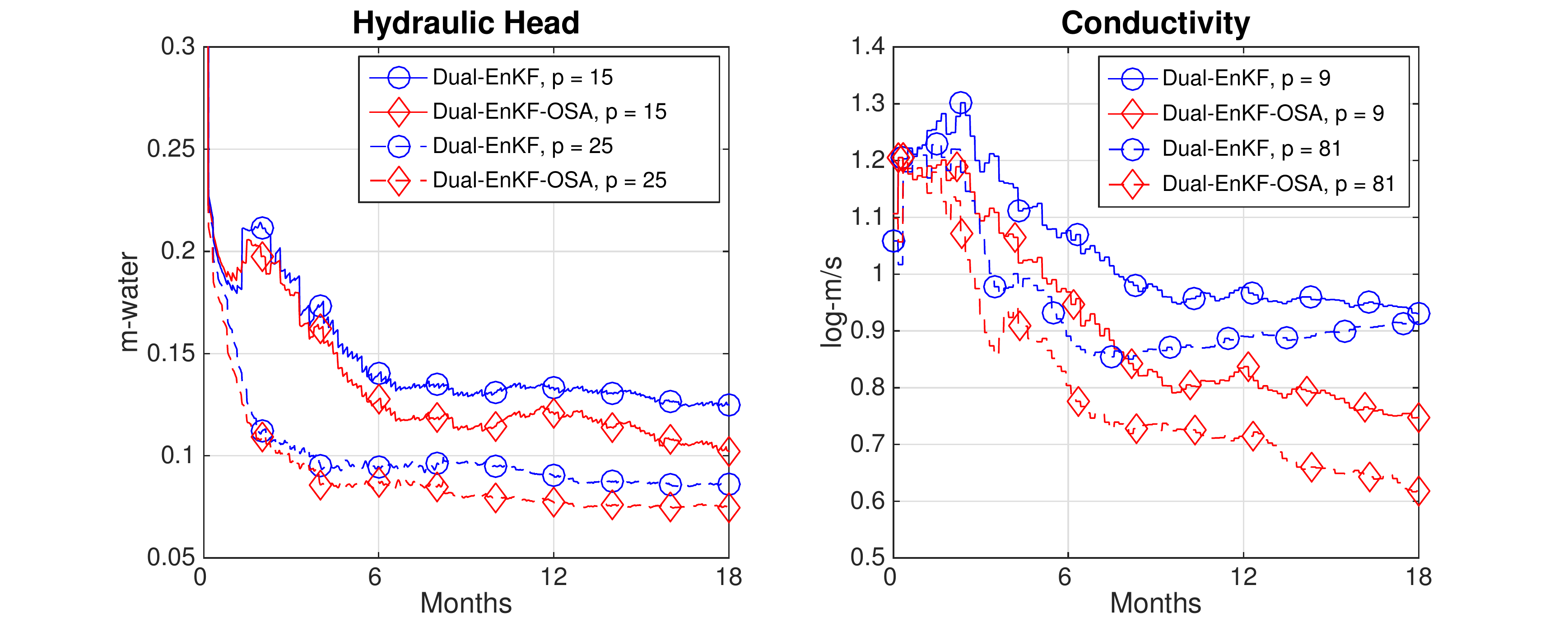} \vskip -0.05cm
\caption{\small Time series of AAE of hydraulic head (left panel) and conductivity (right panel) using the Dual-EnKF and Dual-EnKF$_{\rm OSA}$ schemes. Results are shown for $4$ scenarios in which assimilation of hydraulic head data are obtained from $9$, $15$, $25$ and $81$ wells (uniformly distributed throughout the aquifer domain) every $5$ days. The four experimental scenarios use $100$ ensemble members and $0.10$ m as the measurement error. The number of wells is denoted by $p$.}
\end{figure}

\textcolor{black}{To further assess the performance of the filters we analyze the spatial patterns of the estimated fields. To do so, we plot and interpret the ensemble mean of the conductivity as it results from the three filters using nine observation wells. We compare the estimated  fields after $18$ months (Figure 8) with the reference conductivity. As can be seen, the Joint- and the Dual-EnKFs exhibit some overshooting in the southern (low conductivity) and central regions of the domain. In contrast, the Dual-EnKF$_{\rm OSA}$ better delineates these regions and further provides reasonable estimates of the low conductivity area in the northwest part of the aquifer.}  

\begin{figure}[h]
\centering \vskip - 0.005cm
\includegraphics[width=14cm]{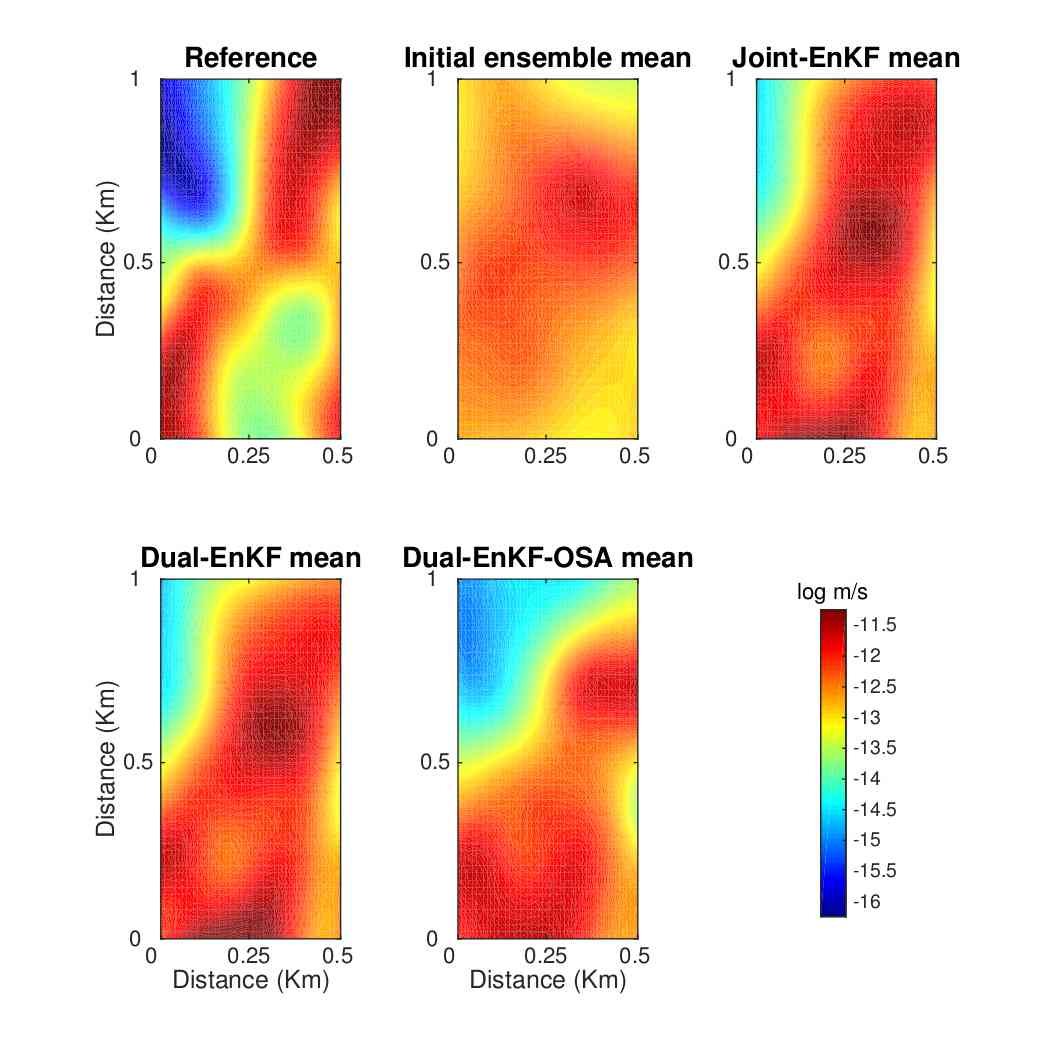} \vskip -0.3cm
\caption{\small Spatial maps of the reference, initial and recovered ensemble means of hydraulic conductivity using the Joint-EnKF, Dual-EnKF, and Dual-EnKF$_{\rm OSA}$ schemes. Results are shown for a scenario in which assimilation of hydraulic head data is obtained from nine wells every five days. This experiment uses $100$ ensemble members and $0.10$ m as the measurement error.}
\end{figure}

\subsection{Sensitivity to measurement errors}
In the last set of sensitivity experiments, we fix the number of wells to nine, the observation frequency to $5$ days, and we use different measurement errors to perturb the observations. We plot the \textcolor{black}{results of nine different observational error scenarios in Figure 9} and compare the conductivity estimates obtained using the Joint-EnKF, Dual-EnKF and the Dual-EnKF$_{\rm OSA}$. In general, the performance of the filters appears to degrade as the observations are perturbed with larger degree of noise. All three filters exhibit similar performances  \textcolor{black}{with large observational error; i.e., 1, 2 and 3 m. This can be expected because larger observational errors decrease the impact of data assimilation, and thus the estimation process is reduced to  a model prediction only.} The plot also suggests that the estimates of the Dual-EnKF$_{\rm OSA}$ with $0.30$ m measurement errors are more accurate than those of the Joint- and the Dual-EnKFs with $0.10$ m error. \textcolor{black}{With $0.10$ m measurement error, the estimate of the Dual-EnKF$_{\rm OSA}$ is also approximately $12\%$ better.  
 }

\begin{figure}[h]
\centering 
\includegraphics[width=10cm]{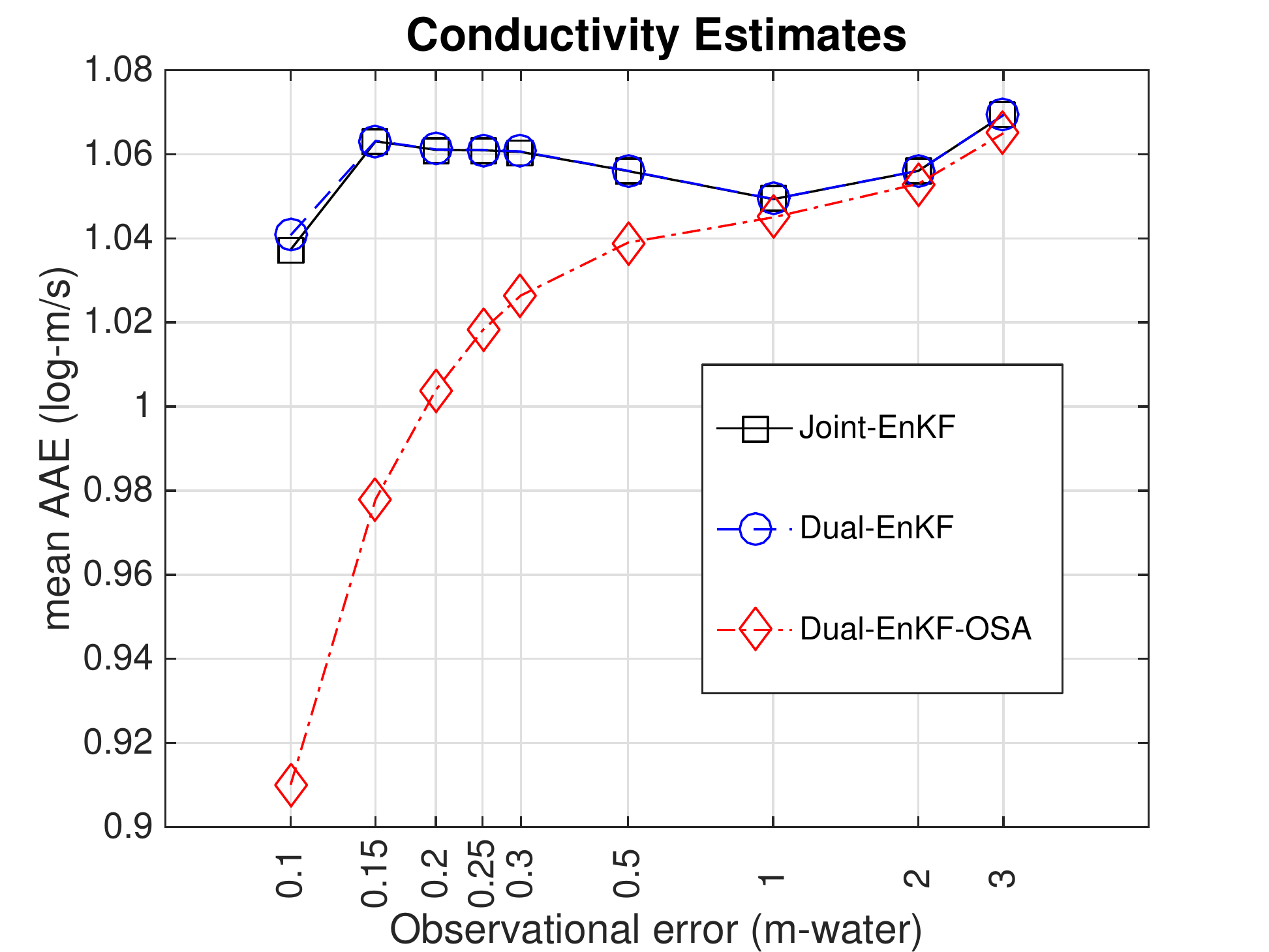}
\caption{\small Mean AAE of the hydraulic conductivity using the Joint-EnKF, Dual-EnKF and Dual-EnKF$_{\rm OSA}$ schemes. Results are shown for $9$ different scenarios in which assimilation of hydraulic head data are obtained from $9$ wells with measurement errors of $0.10$, $0.15$, $0.20$, $0.25$, $0.3$, $0.5$, $1$, $2$ and $3$ m. The four experimental scenarios use $100$ ensemble members and $5$ days as observation frequency.}
\end{figure}

\textcolor{black}{Finally, we investigated the time-evolution of the ensemble variance of the conductivity estimates as they result from  the Dual-EnKF and the Dual-EnKF$_{\rm OSA}$ with $0.10$ m-water measurement noise. Spatially, the ensemble variance maps provide insight about the uncertainty reduction due to data assimilation. The initial map (left panel, Figure 10) exhibits zero variance at the sampled two locations and increasing variance away from these locations. The ensemble spread of conductivity field from the two filters (right panels, Figure 10) after 6 and 18 months is quite small and comparable. The Dual-EnKF$_{\rm OSA}$, however, tends to maintain a larger variance at the edges than  the Dual-EnKF, which in turn increases the impact of the observations.}  


\subsection{Further assessments of the Dual-EnKF$_{\rm OSA}$ scheme}
\textcolor{black}{To further assess the system performance in terms of parameters retrieval, we have 
integrated the model in prediction mode (without assimilation) for an additional period of 18 months starting from the end of the assimilation period. 
 We plot in Figure 11, using the final estimates of the conductivity as they result  from the three filters (after 18 months), the time evolution of the hydraulic head at the control well (CW). The ensemble size is set to $100$, observation frequency is 1 day, number of data wells is 25 and measurement noise is 0.5 m. The reference head trajectory at the CW  decreases from $17.5$ to $16.9$ m in the first 2 years, and then slightly increases to $17.2$ m in the remaining simulation year. The forecast ensemble members of the Joint-EnKF at this CW fail to capture to reference trajectory of the model. This could be due to the large measurement noise imposed on the head data. The Dual-EnKF performs slightly better and predicts hydraulic head values that are closer to the reference solution. The performance of the Dual-EnKF$_{\rm OSA}$, as  shown, is the closest to the reference head trajectory and moreover, one of the forecast  ensemble  members successfully captures the true head evolution. Similar verification was also conducted at other locations in the aquifer and all resulted in similar results.}    
  

Finally, in order to demonstrate that our results are statistically robust, $10$ other test cases with different reference conductivity and heterogeneous recharge maps were investigated. In each of these cases, we sampled the reference fields by varying the variogram parameters, such as variance, $x$ and $y$ ranges, etc. The pumping rates and the initial head configuration among the cases were also altered. For all $10$ test cases, we fixed the ensemble size to $100$ and used data from nine observation wells every $3$ days. We set the measurement error to $0.10$ m. \textcolor{black}{We plot the resulting conductivity estimates (mean AAE) from each case in Figure 12.} The estimates of the three filters, as shown on the plot, give a statistical evidence that the proposed scheme always provides  more accurate estimates than the Joint-/Dual-EnKF and is  more robust to changing dynamics and experimental setups. Similar results were obtained for the hydraulic head estimates. Averaging over all test cases, \textcolor{black}{the proposed scheme provides about  17\% more accurate estimates in term of AAE than the standard Joint- and Dual-EnKFs.}     

\begin{figure}[h]
\centering \vskip -.3cm 
\includegraphics[width=13cm]{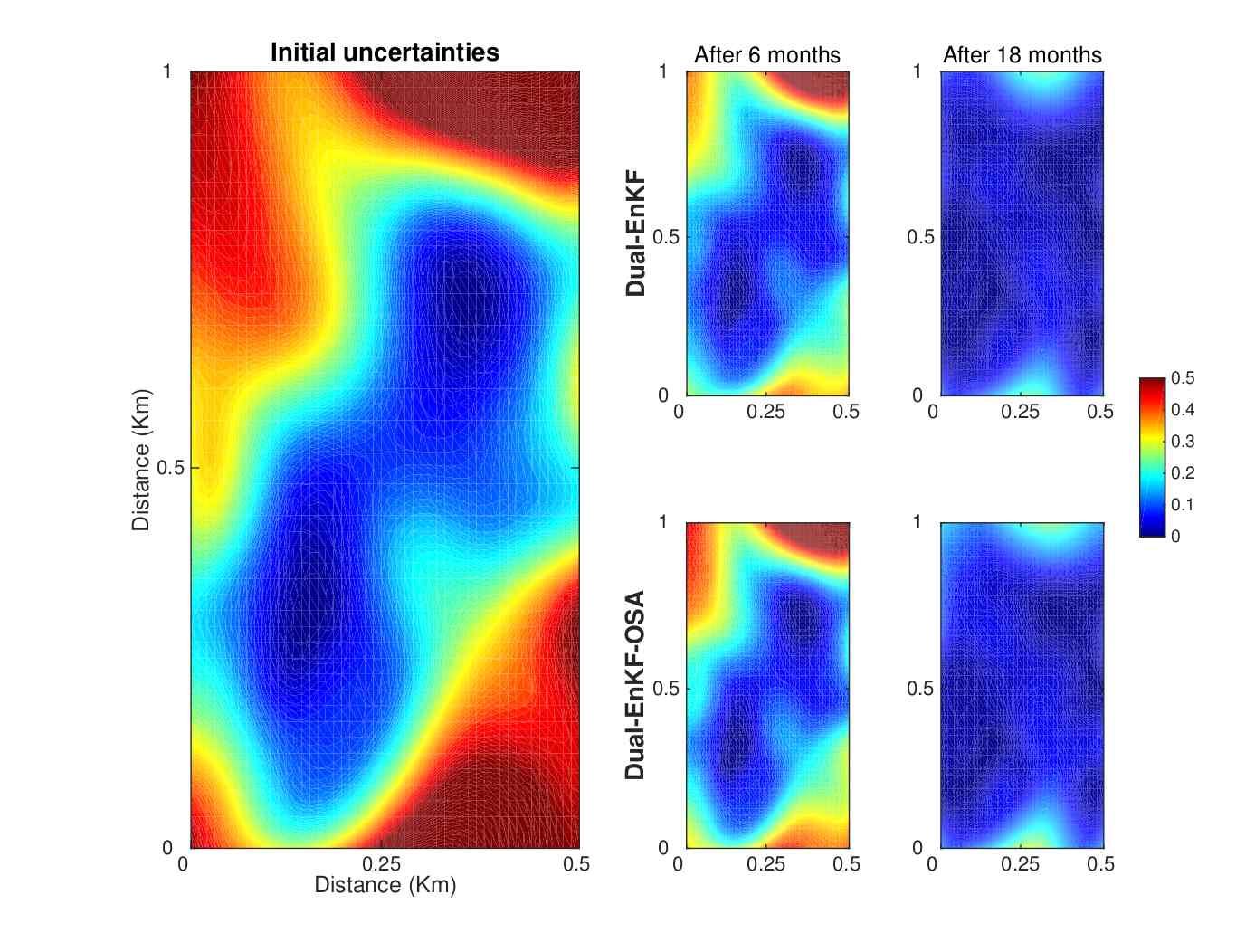} \vskip  -.9cm
\caption{\small Left panel: Ensemble variance map of the initial conductivity field. Right sub-panels: Ensemble variance maps of estimated conductivity after $6$ and $18$ month assimilation periods using the Dual-EnKF and the proposed Dual-EnKF$_{\rm OSA}$ schemes. These results are obtained with $100$ members, $5$ days of observation frequency, $9$ observation wells, and $0.10$ m as measurement noise.}
\end{figure}

\vspace{.5cm}
\begin{figure}[h]
\centering \vskip -.01cm 
\includegraphics[width=14cm]{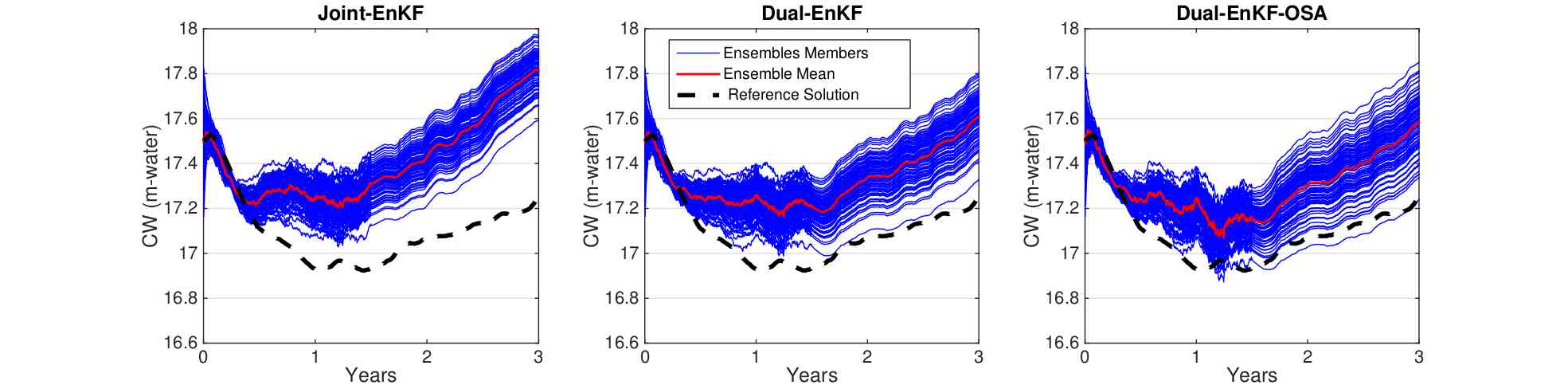} 
\caption{\small Reference (dashed) and predicted (solid) hydraulic head evolution at the control well: CW. Results are obtained using the Joint-EnKF, Dual-EnKF and the Dual-EnKF$_{\rm OSA}$ schemes with $100$ members, $1$ day as observation frequency, $25$ observation wells, and $0.50$ m of measurement noise. The last 18 months are purely based on the forecast model prediction with no assimilation of data.}
\end{figure}

\begin{figure}[h]
\centering 
\includegraphics[width=13cm]{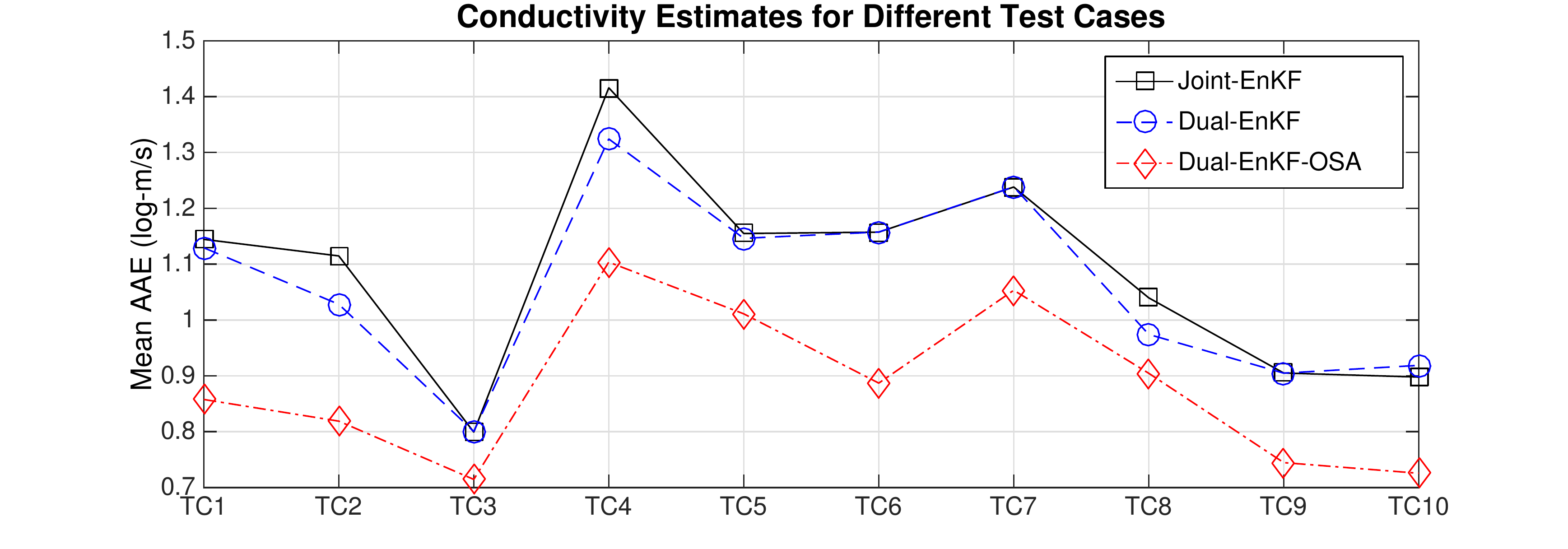} 
\caption{\small Performance of the Joint-/Dual-EnKF and the proposed Dual-EnKF$_{\rm OSA}$ schemes in $10$ different test cases (TC1, TC2, ...). Mean AAE of the conductivity estimates are displayed. These results are obtained with $100$ members, $3$ days of observation frequency, $9$ observation wells, and $0.10$ m as measurement noise.}
\end{figure}

\vspace*{.02cm} 
\section{Conclusion} 
We presented a one-step-ahead smoothing based dual ensemble Kalman filter (Dual-EnKF$_{\rm OSA}$) for state-parameter estimation of subsurface groundwater flow models. The Dual-EnKF$_{\rm OSA}$ is derived using a Bayesian probabilistic formulation combined with two classical stochastic sampling properties. It differs from the standard Joint-EnKF and Dual-EnKF in  the fact that the order of the time-update step of the state (forecast by the model) and the measurement-update step (correction by the incoming observations) is inverted. {\color{black}Compared with the Dual-EnKF, this introduces a smoothing step to the state by future observations, which seems to provide the model, at the time of forecasting, with better and rather physically-consistent state and parameters ensembles.} 
 
We tested the proposed Dual-EnKF$_{\rm OSA}$  on a conceptual groundwater flow model in which we estimated the hydraulic head and spatially variable conductivity parameters. We conducted a  number of sensitivity experiments to evaluate the accuracy and the robustness of the proposed  scheme and to compare its performance against those of the standard Joint  and Dual EnKFs. The  experimental results suggest that the Dual-EnKF$_{\rm OSA}$ is more robust, successfully estimating  the hydraulic head and the conductivity field under different modeling scenarios. Sensitivity analyses demonstrate that with the assimilation of more observations, the Dual-EnKF$_{\rm OSA}$  becomes more effective and significantly outperforms the standard Joint- and Dual-EnKF schemes. In addition, when using a sparse observation network in the aquifer domain, the accuracy of the Dual-EnKF$_{\rm OSA}$ estimates is better preserved, unlike the Dual-EnKF, which seems to be more sensitive to the number of hydraulic wells. Moreover, the Dual-EnKF$_{\rm OSA}$ results are shown to be more robust against observation noise. On average, the Dual-EnKF$_{\rm OSA}$ scheme leads to around 10\% more accurate state and parameters solutions than those resulting from the standard Joint- and Dual-EnKFs.  

The proposed scheme is easy to implement and only requires minimal modifications to a standard EnKF code. It is further computationally feasible, requiring only a marginal increase in the computational cost compared to the Dual-EnKF. This scheme should therefore be beneficial to the hydrology community given its consistency, high accuracy, and robustness to changing modeling conditions. It could serve as an efficient estimation tool for real-world problems, such as groundwater, contaminant transport and reservoir monitoring, in which the available data are often sparse and noisy. Potential future research includes testing the Dual-EnKF$_{\rm OSA}$ with realistic large-scale groundwater, contaminant transport and reservoir monitoring problems. Furthermore, integrating the proposed state-parameter estimation scheme with other iterative and hybrid ensemble approaches may be a promising direction for further improvements.

\section*{\textcolor{black}{Appendix}}
\textcolor{black}{ We show here that the samples, $\widetilde{\bf x}_n^{(m)} ({\bf x}_{n-1} , \theta)$, given in (\ref{sample-a-poetserior-pdf-given-x-3}), are drawn from the {\sl a posteriori} transition pdf, $p({\bf x}_n|  {\bf x}_{n-1}, \theta, {\bf y}_{n})$. Lets start by showing how Eqs. (\ref{sample-a-poetserior-pdf-given-x-1})-(\ref{sample-a-poetserior-pdf-given-x-2}) are obtained.  According to (\ref{eq-trans-MC-z}), on can  show that the members, $\widetilde{\xi}_n^{(m)} ({\bf x}_{n-1} , \theta)$, given by (\ref{sample-a-poetserior-pdf-given-x-1}), are samples  from the transition pdf, $ p({\bf x}_n|{\bf x}_{n-1},\theta) = {\cal N}_{{\bf x}_n} ({\cal M}_{n-1}({\bf x}_{n-1} , \theta) , {\bf Q}_{n-1}  )$. Furthermore, one may use Property 1 
 in (\ref{eq-smoth-likliho-z}), which is recalled here,  
\begin{equation}
\label{eq-smoth-likliho-z-bis}
p({\bf y}_n | {\bf x}_{n-1}, \theta ) = \int \underbrace{ p({\bf y}_n | {\bf x}_n)}_{{\cal N}_{{\bf y}_n}({\bf H}_n {\bf x}_n , {\bf R}_n) } 
\hspace{.2cm} 
\underbrace{ p({\bf x}_n | {\bf x}_{n-1} , \theta)}_{ \approx {\left\{\widetilde{\xi}_n^{(m)} ({\bf x}_{n-1} , \theta)\right\}}_{m=1}^{N_e}  } d{\bf x}_n, 
\end{equation} 
to obtain the members, $\widetilde{\bf y}_n^{(m)} ({\bf x}_{n-1} , \theta)$, given by (\ref{sample-a-poetserior-pdf-given-x-2}); such members are, indeed,  samples from $p({\bf y}_n | {\bf x}_{n-1}, \theta )$. }

\textcolor{black}{ Now, using the samples $\widetilde{\xi}_n^{(m)} ({\bf x}_{n-1} , \theta)$ of $p({\bf x}_n|{\bf x}_{n-1},\theta) =  p({\bf x}_n|{\bf x}_{n-1},\theta, {\bf y}_{0:n-1})$ and the samples $\widetilde{\bf y}_n^{(m)} ({\bf x}_{n-1} , \theta)$ of $p({\bf y}_n | {\bf x}_{n-1}, \theta ) = p({\bf y}_n | {\bf x}_{n-1}, \theta, {\bf y}_{0:n-1})$, one can apply Property 2 
 to  the joint pdf, $p({\bf x}_n, {\bf y}_n | {\bf x}_{n-1}, \theta, {\bf y}_{0:n-1})$,  assuming it Gaussian, to show that  the samples  $\widetilde{\bf x}_n^{(m)} ({\bf x}_{n-1} , \theta)$, given in (\ref{sample-a-poetserior-pdf-given-x-3}), are drawn from the {\sl a posteriori} transition pdf, $p({\bf x}_n|  {\bf x}_{n-1}, \theta, {\bf y}_{0:n}) = p({\bf x}_n|  {\bf x}_{n-1}, \theta, {\bf y}_{n})$. }
 
\section*{acknowledgments}
Research reported in this publication was supported by King Abdullah University of Science and Technology (KAUST).   Datasets and codes of the assimilation experiments 
  can be accessed by contacting the second author: \texttt{mohamad.gharamti@nersc.no}. 

\bibliographystyle{IEEEtran} 
\bibliography{IEEEabrv,sample_bib_08}

\begin{thebibliography}{10}
\providecommand{\url}[1]{#1}
\csname url@samestyle\endcsname
\providecommand{\newblock}{\relax}
\providecommand{\bibinfo}[2]{#2}
\providecommand{\BIBentrySTDinterwordspacing}{\spaceskip=0pt\relax}
\providecommand{\BIBentryALTinterwordstretchfactor}{4}
\providecommand{\BIBentryALTinterwordspacing}{\spaceskip=\fontdimen2\font plus
\BIBentryALTinterwordstretchfactor\fontdimen3\font minus
  \fontdimen4\font\relax}
\providecommand{\BIBforeignlanguage}[2]{{%
\expandafter\ifx\csname l@#1\endcsname\relax
\typeout{** WARNING: IEEEtran.bst: No hyphenation pattern has been}%
\typeout{** loaded for the language `#1'. Using the pattern for}%
\typeout{** the default language instead.}%
\else
\language=\csname l@#1\endcsname
\fi
#2}}
\providecommand{\BIBdecl}{\relax}
\BIBdecl

\bibitem{Samuel2014}
J.~Samuel, P.~Coulibaly, G.~Dumedah, and H.~Moradkhani, ``Assessing model state
  and forecasts variation in hydrologic data assimilation,'' \emph{Journal of
  Hydrology}, 2014.

\bibitem{Chen}
Y.~Chen and D.~Zhang, ``Data assimilation for transient flow in geologic
  formations via ensemble kalman filter,'' \emph{Advances in Water Resources},
  vol.~29, no.~8, pp. 1107--1122, 2006.

\bibitem{Franssen2}
H.~Hendricks~Franssen and W.~Kinzelbach, ``Real-time groundwater flow modeling
  with the ensemble kalman filter: Joint estimation of states and parameters
  and the filter inbreeding problem,'' \emph{Water Resources Research},
  vol.~44, no.~9, 2008.

\bibitem{Gharamti2}
M.~Gharamti and I.~Hoteit, ``Complex step-based low-rank extended kalman
  filtering for state-parameter estimation in subsurface transport models,''
  \emph{Journal of Hydrology}, vol. 509, pp. 588--600, 2014.

\bibitem{Gharamti2014Hybrid}
M.~Gharamti, J.~Valstar, and I.~Hoteit, ``An adaptive hybrid enkf-oi scheme for
  efficient state-parameter estimation of reactive contaminant transport
  models,'' \emph{Advances in Water Resources}, vol.~71, pp. 1--15, 2014.

\bibitem{Vrugt2003}
J.~A. Vrugt, H.~V. Gupta, W.~Bouten, and S.~Sorooshian, ``A shuffled complex
  evolution metropolis algorithm for optimization and uncertainty assessment of
  hydrologic model parameters,'' \emph{Water Resources Research}, vol.~39,
  no.~8, 2003.

\bibitem{Valstar2004}
J.~R. Valstar, D.~B. McLaughlin, C.~Te~Stroet, and F.~C. van Geer, ``A
  representer-based inverse method for groundwater flow and transport
  applications,'' \emph{Water Resources Research}, vol.~40, no.~5, 2004.

\bibitem{Alcolea2006}
A.~Alcolea, J.~Carrera, and A.~Medina, ``Pilot points method incorporating
  prior information for solving the groundwater flow inverse problem,''
  \emph{Advances in water resources}, vol.~29, no.~11, pp. 1678--1689, 2006.

\bibitem{Feyen2007}
L.~Feyen, J.~A. Vrugt, B.~{\'O}. Nuall{\'a}in, J.~van~der Knijff, and
  A.~De~Roo, ``Parameter optimisation and uncertainty assessment for
  large-scale streamflow simulation with the lisflood model,'' \emph{Journal of
  Hydrology}, vol. 332, no.~3, pp. 276--289, 2007.

\bibitem{Franssen1}
H.~Franssen and W.~Kinzelbach, ``Ensemble kalman filtering versus sequential
  self-calibration for inverse modelling of dynamic groundwater flow systems,''
  \emph{Journal of hydrology}, vol. 365, no.~3, pp. 261--274, 2009.

\bibitem{Elsheikh2013}
A.~H. Elsheikh, M.~F. Wheeler, and I.~Hoteit, ``An iterative stochastic
  ensemble method for parameter estimation of subsurface flow models,''
  \emph{Journal of Computational Physics}, vol. 242, pp. 696--714, 2013.

\bibitem{Chang}
S.-Y. Chang, T.~Chowhan, and S.~Latif, ``State and parameter estimation with an
  sir particle filter in a three-dimensional groundwater pollutant transport
  model,'' \emph{Journal of Environmental Engineering}, vol. 138, no.~11, pp.
  1114--1121, 2012.

\bibitem{livredoucet}
A.~Doucet, N.~de~Freitas, and N.~Gordon, Eds., \emph{Sequential Monte Carlo
  Methods in Practice}, ser. Statistics for Engineering and Information
  Science.\hskip 1em plus 0.5em minus 0.4em\relax New York: Springer Verlag,
  2001.

\bibitem{Moradkhani}
H.~Moradkhani, K.-L. Hsu, H.~Gupta, and S.~Sorooshian, ``Uncertainty assessment
  of hydrologic model states and parameters: Sequential data assimilation using
  the particle filter,'' \emph{Water Resources Research}, vol.~41, no.~5, 2005.

\bibitem{Montzka}
C.~Montzka, H.~Moradkhani, L.~Weiherm{\"u}ller, H.-J.~H. Franssen, M.~Canty,
  and H.~Vereecken, ``Hydraulic parameter estimation by remotely-sensed top
  soil moisture observations with the particle filter,'' \emph{Journal of
  hydrology}, vol. 399, no.~3, pp. 410--421, 2011.

\bibitem{Reichle2002}
R.~H. Reichle, D.~B. McLaughlin, and D.~Entekhabi, ``Hydrologic data
  assimilation with the ensemble kalman filter,'' \emph{Monthly Weather
  Review}, vol. 130, no.~1, pp. 103--114, 2002.

\bibitem{Vrugt2006}
J.~A. Vrugt, H.~V. Gupta, B.~Nuall{\'a}in, and W.~Bouten, ``Real-time data
  assimilation for operational ensemble streamflow forecasting,'' \emph{Journal
  of Hydrometeorology}, vol.~7, no.~3, pp. 548--565, 2006.

\bibitem{Zhou2011}
H.~Zhou, J.~J. G{\'o}mez-Hern{\'a}ndez, H.-J. Hendricks~Franssen, and L.~Li,
  ``An approach to handling non-gaussianity of parameters and state variables
  in ensemble kalman filtering,'' \emph{Advances in Water Resources}, vol.~34,
  no.~7, pp. 844--864, 2011.

\bibitem{Gharamti2013}
M.~E. Gharamti, I.~Hoteit, and J.~Valstar, ``Dual states estimation of a
  subsurface flow-transport coupled model using ensemble kalman filtering,''
  \emph{Advances in Water Resources}, vol.~60, pp. 75--88, 2013.

\bibitem{Mclaughlin2002}
D.~McLaughlin, ``An integrated approach to hydrologic data assimilation:
  interpolation, smoothing, and filtering,'' \emph{Advances in Water
  Resources}, vol.~25, no.~8, pp. 1275--1286, 2002.

\bibitem{Wan1999}
E.~A. Wan, R.~Van Der~Merwe, and A.~T. Nelson, ``Dual estimation and the
  unscented transformation.'' in \emph{NIPS}.\hskip 1em plus 0.5em minus
  0.4em\relax Citeseer, 1999, pp. 666--672.

\bibitem{Nevadal}
G.~N{\ae}vdal, L.~M. Johnsen, S.~I. Aanonsen, E.~H. Vefring \emph{et~al.},
  ``Reservoir monitoring and continuous model updating using ensemble kalman
  filter,'' \emph{SPE journal}, vol.~10, no.~01, pp. 66--74, 2005.

\bibitem{Li2012}
L.~Li, H.~Zhou, J.~J. G{\'o}mez-Hern{\'a}ndez, and H.-J. Hendricks~Franssen,
  ``Jointly mapping hydraulic conductivity and porosity by assimilating
  concentration data via ensemble kalman filter,'' \emph{Journal of Hydrology},
  vol. 428, pp. 152--169, 2012.

\bibitem{Moradkhani2005}
H.~Moradkhani, S.~Sorooshian, H.~V. Gupta, and P.~R. Houser, ``Dual
  state--parameter estimation of hydrological models using ensemble kalman
  filter,'' \emph{Advances in Water Resources}, vol.~28, no.~2, pp. 135--147,
  2005.

\bibitem{Gharamti2014}
M.~Gharamti, A.~Kadoura, J.~Valstar, S.~Sun, and I.~Hoteit, ``Constraining a
  compositional flow model with flow-chemical data using an ensemble-based
  kalman filter,'' \emph{Water Resources Research}, 2014.

\bibitem{livreRobert-bayesianchoice-endEdit}
C.~Robert, \emph{{T}he {B}ayesian {C}hoice: {F}rom {D}ecision-{T}heoretic
  {F}oundations to {C}omputational {I}mplementation}.\hskip 1em plus 0.5em
  minus 0.4em\relax New York: Springer Science \& Business Media, 2007.

\bibitem{y-hoffman-et-al-1991}
Y.~Hoffman and E.~Ribak, ``{C}onstrained realizations of {G}aussian fields - a
  simple algorithm,'' \emph{The Astrophysical Journal}, vol. 380, no. 491, pp.
  L5--L8, 1991.

\bibitem{WenChen2006}
X.~H. Wen and W.~H. Chen, ``Real-time reservoir updating using ensemble kalman
  filter: The confirming approach,'' \emph{Society of Petroleum Engineering},
  vol.~11, pp. 431--442, 2007.

\bibitem{desbouvries-et-al-2011}
F.~Desbouvries, Y.~Petetin, and B.~Ait-El-Fquih, ``{D}irect, {P}rediction- and
  {S}moothing-based {K}alman and {P}article {F}ilter {A}lgorithms,''
  \emph{Signal Processing}, vol.~91, no.~8, pp. 2064--2077, 2011.

\bibitem{w-lee-and-c-farmer-2014}
W.~Lee and C.~Farmer, ``{D}ata {A}ssimilation by {C}onditioning of {D}riving
  {N}oise on {F}uture {O}bservations,'' \emph{IEEE Transactions on Signal
  Processing}, vol.~62, no.~15, pp. 3887--3896, 2014.

\bibitem{Bailey2010}
R.~Bailey and D.~Ba{\`u}, ``Ensemble smoother assimilation of hydraulic head
  and return flow data to estimate hydraulic conductivity distribution,''
  \emph{Water Resources Research}, vol.~46, no.~12, 2010.

\bibitem{Gomez1993}
J.~J. G{\'o}mez-Hern{\'a}ndez and A.~G. Journel, ``Joint sequential simulation
  of multigaussian fields,'' in \emph{Geostatistics TroiaÕ92}.\hskip 1em plus
  0.5em minus 0.4em\relax Springer, 1993, pp. 85--94.

\end{thebibliography}

\end{document}